\documentclass[prepreint,twocolumn,trackchanges]{aastex62}
\hypersetup{linkcolor={blue},citecolor={blue}}
\usepackage{cprotect}
\received{\today}
\revised{\today}
\accepted{\today}
\submitjournal{ApJ}

\shorttitle{SN ejecta with a central power source}
\shortauthors{Suzuki \& Maeda}

\begin{document}
\title{2D radiation-hydrodynamic simulations of supernova ejecta with a central power source}

\correspondingauthor{Akihiro Suzuki}
\email{akihiro.suzuki@nao.ac.jp}

\author[0000-0002-7043-6112]{Akihiro Suzuki}
\affil{Division of Science, National Astronomical Observatory of Japan, 2-21-1 Osawa, Mitaka, Tokyo 181-8588, Japan}

\author[0000-0003-2611-7269]{Keiichi Maeda}
\affil{Department of Astronomy, Kyoto University, Kitashirakawa-Oiwake-cho, Sakyo-ku, Kyoto 606-8502, Japan}



\begin{abstract}
We present the results of two-dimensional radiation-hydrodynamic simulations of expanding supernova ejecta with a central energy source. 
As suggested in previous multi-dimensional hydrodynamic simulations, a sufficiently powerful central energy source can blow away the expanding supernova ejecta, leading to efficient mixing of stratified layers in the ejecta. 
We assume that the energy injection is realized in the form of non-thermal radiation from the wind nebula embedded at the center of the ejecta. 
We found that the multi-dimensional mixing in the ejecta assists the injected non-thermal radiation escaping from the ejecta. 
When the non-thermal radiation is absorbed by the ejecta, it is converted into bright thermal radiation or is consumed as the kinetic energy of the supernova ejecta. 
We found that central energy sources with the injection timescale similar to the photon diffusion timescale realize an efficient conversion of the injected energy into thermal radiation. 
On the other hand, a rapid energy injection ends up accelerating the ejecta rather than giving rise to bright thermal emission. 
This remarkable difference potentially explains the diversity of energetic supernovae including broad-lined Ic and superluminous supernovae. 
\end{abstract}



\section{INTRODUCTION\label{intro}}
Transient surveys in the last few decades have revealed diverse classes of stellar explosions resulting from the gravitational collapse of massive stars, i.e., core-collapse supernovae (CCSNe). 
Superluminous supernovae (SLSNe) are among the most important discoveries made by such unbiased transient surveys \citep{2011ApJ...743..114C,2011Natur.474..487Q,2012Sci...337..927G}. 
These rare, but extremely bright explosions originate from massive stars in special conditions (see, e.g., \citealt{2018SSRv..214...59M,2019ARA&A..57..305G} for reviews, but see \citealt{2020Sci...367..415J}). Although they are intriguing objects potentially found even in star-forming galaxies in the high-z universe \citep{2017MNRAS.470.4241P,2018ApJ...854...37S,2019ApJS..241...17C,2019ApJS..241...16M}, the power source of their bright thermal emission is still extensively debated. 
SLSNe are divided into two spectral classes, Type-I and Type-II SLSNe, according to the absence and the presence of hydrogen spectral features. 
Type-II SLSNe are considered to be extreme cases of SNe colliding with their massive hydrogen-rich circumstellar media (CSM). 
The absence of hydrogen and helium features in the spectra of Type-I SLSNe (referred to as SLSNe-I hereafter) indicates that their progenitors are the bare carbon-oxygen core of massive stars. 
Despite extensive studies since their discovery, the origin of SLSNe-I remains unsolved.

The peak magnitudes of SLSNe-I distribute around $M\sim -21$ \citep{2018ApJ...860..100D,2018ApJ...852...81L,2019MNRAS.487.2215A}. 
The extreme brightness of SLSNe-I does not reconcile with the standard emission mechanism of CCSNe of Type-I, i.e. the radioactive decay of freshly synthesized $^{56}$Ni at the moment of the explosion \citep{1969ApJ...157..623C,1980ApJ...237..541A,1982ApJ...253..785A}, because the required amount of $^{56}$Ni reaches a few $M_\odot$ or more. 
Currently, the collapse of very massive stars with the initial mass exceeding $\sim 130M_\odot$, i.e., the pair-instability SN \citep{1967PhRvL..18..379B,1967ApJ...148..803R,2002ApJ...567..532H}, is the only hypothesized channel to produce such a large amount of $^{56}$Ni. 
However, their expected properties are in disagreement with observations in terms of the light curve \citep[e.g.,][]{2013Natur.502..346N} and spectral evolution \citep[e.g.,][]{2012MNRAS.426L..76D,2016MNRAS.455.3207J}. 

Alternatively, extra power sources have been paid a lot of attention. 
Currently, two leading scenarios are extensively discussed; the CSM interaction \citep[e.g.,][]{2011ApJ...729L...6C,2012ApJ...757..178G,2013MNRAS.428.1020M} and the central engine (\citealt{2010ApJ...717..245K,2010ApJ...719L.204W}; see also \citealt{2007ApJ...666.1069M}) scenarios. 
In the former scenario, the SN is surrounded by a massive hydrogen-poor CSM. 
The collision between the SN ejecta and the CSM converts the kinetic energy of the expanding ejecta into thermal radiation. 
This scenario also includes the pulsational pair-instability model, in which a very massive star experiences eruptive mass-loss due to the pair-instability before the terminal explosion \citep{2007Natur.450..390W}. 
The latter scenario considers a central compact object, which is often presumed to be a magnetized proto-neutron star with an initial spin period of $\sim 1$ ms, and its rotational energy loss.  
Another proposed candidate is black hole accretion disks \citep{2013ApJ...772...30D}, although its efficiency to extract the accretion energy and to power the SN ejecta is recently questioned \citep{2018ApJ...867..113M}. 
A number of simple light-curve models based on both the CSM and central engine scenarios are proposed and successfully explain photometric observations of SLSNe-I \citep{2012ApJ...746..121C,2013ApJ...773...76C,2013ApJ...770..128I,2015MNRAS.452.3869N,2017ApJ...850...55N,2015ApJ...807..147W,2016ApJ...821...22W}, which makes it difficult to distinguish one scenario from the other only by photometric observations. 
Therefore, additional information from follow-up observations in multi-wavelengths would be necessary. 
Spectral similarities of SLSNe-I to broad-lined Type-Ic SNe (SNe Ic-BL) \citep[e.g.,][]{2010ApJ...724L..16P,2017ApJ...845...85L}, which are sometimes associated with gamma-ray bursts, and the recent detection of bright radio source associated with an almost 10 years-old SLSN, PTF10hgi \citep{2019ApJ...886...24L}, suggest that some SLSNe-I certainly harbor a powerful central engine. 
In this paper, we focus on the central engine scenario. 

How exactly the rotational energy of the central compact object is extracted and deposited in the surrounding SN ejecta is unclear. 
One possibility is that the central compact object produces a magnetized wind, which pushes the surrounding gas in a similar way to Galactic pulsar wind nebulae, such as the Crab nebula. 
In such a case, non-thermal emission from the wind nebula heats the SN ejecta. 
The continuous energy input would influence the density and ionization structure of the SN ejecta and leads to the appearance of highly ionized elements, such as \ion{O}{2} and \ion{O}{3}, in the optical spectra. 
Investigating observational imprints of the putative central engine embedded in SLSNe in both theoretical and observational ways is therefore important for distinguishing the central engine scenario from competing ones.

One of the important ingredients in the energy transport in the SN ejecta is multi-dimensional effects. 
In previous studies, we have performed hydrodynamic simulations of SN ejecta with a central energy source in 2D \citep{2017MNRAS.466.2633S} and 3D \citep{2019ApJ...880..150S}. 
These simulations have revealed that the impact of the central energy injection on the SN ejecta is not negligible at all (see also \citealt{2016ApJ...832...73C,2020ApJ...893...99C,2017ApJ...845..139B}). 
In particular, the acceleration of the outer ejecta to mildly relativistic speeds may produce bright synchrotron and inverse Compton emission, which can be probed by multi-wavelength follow-up observations and potentially used for the model discrimination  \citep{2019ApJ...870...38S}. 
Although a lot of important effects are clarified by multi-dimensional studies of central engine-driven SN ejecta, previous studies have been based on hydrodynamic simulations without radiative transfer effects. 
In the central engine model for SLSNe, however, the timescale of the energy injection needs to be comparable to the photon diffusion timescale throughout the SN ejecta in order to power the bright thermal emission. 
This means that radiative transfer effects in the ejecta are an indispensable ingredient for modeling SLSNe. 

In this study, we perform first two-dimensional radiation-hydrodynamic simulations of expanding SN ejecta with a central energy source under a gray approximation. 
This paper is organized as follows. 
In section \ref{sec:numerical_setups}, we outline the numerical setups of our simulations. 
Section \ref{sec:results} presents the results of the simulations with a special focus on the multi-dimensional and radiative transfer effects. 
We discuss some implications for the observed properties of engine-driven SN in Section \ref{sec:discussion}. 
Finally, we conclude this paper in Section \ref{sec:conclusions}. 
In the following, we adopt the unit $c=1$ unless otherwise explicitly stated.

\section{Numerical procedures and setups\label{sec:numerical_setups}}
This section describes the setups of our numerical simulations. 
We use the radiation-hydrodynamic simulation code developed by one of the authors (AS). 
The same basic strategy as our previous work \citep{2016ApJ...825...92S,2019ApJ...887..249S} is employed to solve the radiation-hydrodynamics equations incorporating special relativity except for the introduction of gamma-ray transport. 
In the following, we briefly describe our numerical approach.  
Although our numerical simulations are performed in cylindrical coordinates $(r,z)$, the following governing equations are expressed in three-dimensional cartesian coordinates for the sake of general applications. 
It is straightforward to include geometric factors in curvilinear coordinates. 
We also note that we use the Einstein summation convention. 

\subsection{Treatment of radiative transfer\label{sec:radiative_transfer}}
In our simulations, the radiation moment equations are solved along with the hydrodynamic equations. 
For the integration of the radiative transfer equation, we use two inertial frames, the laboratory and comoving frames. We denote physical variables in the laboratory and comoving frames by letters with and without overbars, e.g., $Q$ and $\bar{Q}$. 

In the central engine scenario for SLSNe, it is often assumed that the energy injection is initially realized by non-thermal emission from the wind nebula at the center of the expanding SN ejecta. 
As long as the SN ejecta is sufficiently opaque to such non-thermal photons, the injected non-thermal photons are immediately converted to thermal energy or photons after propagating at distances comparable to the mean free path of non-thermal photons. 
At later epochs, however, the SN ejecta is gradually becoming transparent, allowing them easily escape into the surrounding space. 
The leakage of non-thermal photons could reduce the energy deposition rate and thus affect light curves \citep[e.g.,][]{2015ApJ...799..107W}, although the expected spectrum and the opacity for non-thermal photons are highly uncertain \citep[e.g.,][]{2013MNRAS.432.3228K}. 
In order to mimic such situations, we divide the radiation into two components, thermal and non-thermal ones.

\subsubsection{Equations}
The hydrodynamic equations govern the temporal evolution of the density $\bar{\rho}$, the velocity $\beta^i$ (normalized by the speed of light), and the gas energy density $\bar{E}_\mathrm{g}$, where the greek index $i$ stands for the spatial components ($i=x,y,z$). 
We solve the standard set of the hydrodynamic equations including special relativistic effects \citep[e.g.,][]{1999LRR.....2....3M}. 
We assume an ideal gas equation of state with an adiabatic index $\gamma=5/3$, for the gas component. 
The gas pressure $\bar{P}_\mathrm{g}$ is therefore given by
\begin{equation}
    \bar{P}_\mathrm{g}=(\gamma-1)\bar{E}_\mathrm{g}.
\end{equation}
On the other hand, radiation moment equations govern the temporal evolution of the frequency-integrated radiation energy density $E_{\mathrm{r},s}$ and flux $F^i_{\mathrm{r},s}$ ($s=$th and nt for the thermal and non-thermal radiation components). 

The radiation energy density $E_{\mathrm{r},s}$ and flux $F^i_{\mathrm{r},s}$ in the laboratory and comoving frames are related with each other by the following formulae,
\begin{equation}
{E}_{\mathrm{r},s}=\Gamma^2\left(\bar{E}_{\mathrm{r},s}+2\beta_i\bar{F}_{\mathrm{r},s}^i+\beta_j\beta_k\bar{P}_{\mathrm{r},s}^{jk}\right),
\label{eq:transformation_Er}
\end{equation}
and
\begin{eqnarray}
\hspace{-2em}
{F}_{\mathrm{r},s}^i&=&\Gamma \bar{F}_{\mathrm{r},s}^i+\Gamma\beta_j \bar{P}_{\mathrm{r},s}^{ij}
\nonumber\\
&&+\Gamma^2\beta^i
\left(\bar{E}_{\mathrm{r},s}+\frac{2\Gamma+1}{\Gamma+1}\beta_j\bar{F}_{\mathrm{r},s}^j+\frac{\Gamma}{\Gamma+1}\beta_j\beta_k\bar{P}_{\mathrm{r},s}^{jk}\right),
\label{eq:transformation_Fr}
\end{eqnarray}
\citep[e.g.,][]{1984oup..book.....M}. 

The radiation energy density and flux evolve according to the following equations,
\begin{equation}
    \frac{\partial E_\mathrm{r,th}}{\partial t}+\frac{\partial F_\mathrm{r,th}^i}{\partial x^i}=G^0_\mathrm{th}+G^0_\mathrm{nt},
\end{equation} 
and
\begin{equation}
    \frac{\partial F_\mathrm{r,th}^i}{\partial t}+\frac{\partial P_\mathrm{r,th}^{ij}}{\partial x^j}=G^i_\mathrm{th}+G^i_\mathrm{nt},
\end{equation}
for the thermal component, and
\begin{equation}
    \frac{\partial E_\mathrm{r,nt}}{\partial t}+\frac{\partial F_\mathrm{r,nt}^i}{\partial x^i}=-G^0_\mathrm{nt},
\end{equation} 
and
\begin{equation}
    \frac{\partial F_\mathrm{r,nt}^i}{\partial t}+\frac{\partial P_\mathrm{r,nt}^{ij}}{\partial x^j}=-G^i_\mathrm{nt},
\end{equation}
for the non-thermal component. 
The former two equations are ordinary moment equations for radiation coupled with gas except for the additional source terms, $G^0_\mathrm{nt}$ and $G^i_\mathrm{nt}$, representing the coupling between the thermal and non-thermal components. 
The latter two equations govern the propagation and the interaction of non-thermal photons. 
The thermal component interacts with gas by the following coupling terms,
\begin{equation}
G^0_\mathrm{th}=\Gamma\bar{\rho}\bar{\kappa_\mathrm{a}}(a_\mathrm{r}\bar{T}_\mathrm{g}^4-\bar{E}_\mathrm{r,th})
-\Gamma\bar{\rho}(\bar{\kappa}_\mathrm{a}+\bar{\kappa}_\mathrm{s})\beta_i\bar{F}_\mathrm{r,th}^i,
\label{eq:G0}
\end{equation}
and
\begin{eqnarray}
G^i_\mathrm{th}&=&
-\bar{\rho}(\bar{\kappa}_\mathrm{a}+\bar{\kappa}_\mathrm{s})\bar{F}_\mathrm{r,th}^i
+\Gamma\bar{\rho}\bar{\kappa}_\mathrm{a}\left(a_\mathrm{r}\bar{T}_\mathrm{g}^4-\bar{E}_\mathrm{r,th}\right)\beta^i
\nonumber\\&&
-\frac{\Gamma^2}{\Gamma+1}\bar{\rho}(\bar{\kappa}_\mathrm{a}+\bar{\kappa}_\mathrm{s})\beta^i\beta_j\bar{F}_\mathrm{r,th}^j.
\label{eq:Gi}
\end{eqnarray}
Here the absorption and scattering opacities are denoted by $\bar{\kappa}_\mathrm{a}$ and $\bar{\kappa}_\mathrm{s}$, respectively. 
We assume that non-thermal photons interact with matter at a constant opacity $\kappa_\mathrm{nt}$. 
When the non-thermal photons are absorbed by gas, they are immediately converted to the thermal counterpart and then the non-thermal radiation energy density and flux decrease at the following rates,
\begin{equation}
    G^0_\mathrm{nt}=\Gamma\bar{\rho}\bar{\kappa}_\mathrm{nt}\bar{E}_\mathrm{r,nt},
\end{equation}
and
\begin{equation}
    G^i_\mathrm{nt}=\Gamma\bar{\rho}\bar{\kappa}_\mathrm{nt}\bar{F}_\mathrm{r,nt}^i.
\end{equation}
We numerically solve these thermal and non-thermal radiation transfer equations in a similar way to our previous work \citep{2019ApJ...887..249S}, which provides detailed numerical procedures for the treatment of the source terms. 
The advection part of the equations (left-hand sides) is integrated by a standard explicit finite volume method with M1 closure \citep{1984JQSRT..31..149L}. 
On the other hand, the source terms of the equations are treated in an implicit way.

As we shall see below, one of the important aspects of the central energy injection is the mixing of material in SN ejecta. 
In order to investigate how inner layers are mixed into outer parts, we consider the transport of elements. 
In practice, we consider several layers with different elemental compositions in the ejecta, whose mass fractions are denoted by $X_l$ (the subscript $l$ denotes a specific layer), and calculate the evolution of the mass fraction distribution by solving the following transport equation,
\begin{equation}
    \frac{\partial(\bar{\rho} X_l)}{\partial t}+\frac{\partial (\bar{\rho} X_l\beta^i)}{\partial x^i}=0,
\end{equation}
along with the equations of hydrodynamics.

\subsubsection{Radiative processes}
We assume that free-free absorption/emission and electron scattering are dominant radiative processes in the SN ejecta and gas is fully ionized, which is a reasonable approximation for early SN evolution. 
The absorption coefficient for free-free absorption is given by
\begin{equation}
\bar{\kappa}_\mathrm{a}=6.64\times 10^{22}\frac{\bar{Z}^3}{\bar{A}^2}\rho \bar{T}_\mathrm{g}^{-7/2}\ \mathrm{cm^2\ g^{-1}},
\label{eq:kappa_a}
\end{equation}
(in cgs units; \citealt{1979rpa..book.....R})
where the average mass and change numbers are set to $\bar{Z}=8$ and $\bar{A}=16$, while the electron scattering coefficient is given by
\begin{equation}
\bar{\kappa}_\mathrm{s}=0.2\ \mathrm{cm^2\ g^{-1}},
\label{eq:kappa_es}
\end{equation}
(\citealt{1979rpa..book.....R}). 

The opacity $\bar{\kappa}_\mathrm{nt}$ for non-thermal photons is highly uncertain, because of unknown non-thermal photon spectra from the embedded wind nebula. 
The opacity can vary widely depending on the  energy of the non-thermal photons \citep[e.g.,][]{2013MNRAS.432.3228K}. 
The opacity for a photon energy around $1$MeV, at which photons interact with gas via Compton scattering, is of the order of $0.1$ cm$^2$ g$^{-1}$. 
Photons with higher energies suffer from less significant Compton scattering and the opacity is instead dominated by pair processes $\sim0.01$ cm$^2$ g$^{-1}$ at $>10$ MeV. 
On the other hand, photons with lower energies ($<10$ keV) are efficiently absorbed via photoelectric absorption and thus the opacity increases by many orders of magnitude with $\nu^{-3}$. 
Several studies have performed light curve fittings for SLSNe-I by simply assuming a constant opacity  and required a wide range of values from $\bar{\kappa}_\mathrm{nt}=0.01$ to $0.8$ cm$^2$ g$^{-1}$ \citep[e.g.,][]{2017ApJ...842...26L,2017ApJ...850...55N}. 

In our simulations, we treat non-thermal radiation as a single radiation component without energy dependence, i.e., gray approximation. 
Therefore, we simply set a constant opacity irrespective of the mean energy of non-thermal photons and its temporal evolution. 
We assume a value comparable to the electron scattering opacity,
\begin{equation}
\bar{\kappa}_\mathrm{nt}=0.1\ \mathrm{cm^2\ g^{-1}}.
\end{equation}

The non-thermal radiation transport plays an important role in the energy deposition and therefore should be treated carefully. 
In particular, our assumption of a single representative opacity for non-thermal photons and the conversion into thermal photons may be too simplified. 
For example, non-thermal photons propagating in the ejecta scatter off electrons many times and gradually deposit their energies rather than the immediate conversion into thermal photons assumed above. 
The injected high-energy photons may produce electron-positron pairs and they may interact with the surrounding gas with different opacity. 
The energy injection may also be accompanied by other high-energy particles rather than non-thermal photons. 
Given that the details of the energy injection and the corresponding opacity are highly unclear for SLSNe-I, however, this simple treatment would be reasonable as a first step. 
As a test problem for the non-thermal radiation transport, we calculate the light curve of a Type-Ia SN in Appendix \ref{sec:gamma_ray}. 
We have confirmed that the light curve of a Type-Ia SN with the well-known ejecta property, power source, and the corresponding non-thermal opacity, can be well reproduced at least around the peak even with our simplified treatment by adopting appropriate non-thermal opacity. 
Therefore, we employ this simplified non-thermal radiation transport in this work, while keeping in mind the possibility of further non-thermal processes at play.

\subsection{Supernova ejecta}
We use the same SN ejecta model as our previous hydrodynamic simulations \citep{2017MNRAS.466.2633S,2019ApJ...880..150S} for the purpose of comparison. 
At the beginning of each simulation ($t=t_0=1000$ s), we assume a freely expanding ejecta with spherical symmetry. 
The radial velocity $v_R$ is given by
\begin{equation}
    v_\mathrm{R}=\frac{R}{t_0},
\end{equation}
with $R=(r^2+z^2)^{1/2}$ for velocities smaller than the maximum velocity $v_\mathrm{max}$. 
We use the capital letter $R$ to denote the 3-dimensional radius, which is distinguished from the radial component $r$ of 2D cylindrical coordinates. 
For the initial density distribution, we assume the commonly used double power-law density profile,
\begin{equation}
    \rho_\mathrm{ej}(t_0,R)=
    \left\{
    \begin{array}{ccl}
        \rho_0\left(\frac{v_\mathrm{R}}{v_\mathrm{br}}\right)^{-\delta}&\mathrm{for}&v\leq v_\mathrm{br},\\
        \rho_0\left(\frac{v_\mathrm{R}}{v_\mathrm{br}}\right)^{-m}&\mathrm{for}&v_\mathrm{br}\leq v\leq v_\mathrm{max},\\
    \end{array}
    \right.
\end{equation}
\citep{1989ApJ...341..867C,1999ApJ...510..379M} with $\delta=1$, $m=10$, and $v_\mathrm{max}=10v_\mathrm{br}$. 
The characteristic velocity $v_\mathrm{br}$ divides the ejecta into the inner and outer components and is expressed in terms of the mass $M_\mathrm{ej}$ and the kinetic energy $E_\mathrm{sn}$ of the SN ejecta,
\begin{equation}
    v_\mathrm{br}=\left(\frac{2f_5E_\mathrm{sn}}{f_3M_\mathrm{ej}}\right)^{1/2},
\end{equation}
where the numerical factor $f_l$ is given by
\begin{equation}
    f_l=\frac{(m-l)(l-\delta)}{m-\delta-(l-\delta)(v_\mathrm{max}/v_\mathrm{br})^{m-l}}.
\end{equation}
We assume $M_\mathrm{ej}=10\ M_\odot$ and $E_\mathrm{sn}=10^{51}$ erg and therfore the characteristic velocity yields $4.5\times 10^8$ cm s$^{-3}\simeq 0.015c$. 
The characteristic density $\rho_0$ at the interface between the inner and the outer parts of the ejecta is given by
\begin{equation}
    \rho_0=\frac{f_3M_\mathrm{ej}}{4\pi v_\mathrm{br}^3t_0^3}.
\end{equation}

The diffusion timescale of photons in the ejecta plays an important role in the temporal evolution of the ejecta and the thermal emission. 
In general, the evolutionary timescale of SN light curves is given by the dynamical timescale equal to the photon diffusion timescale at the epoch.  
For the SN ejecta adopted here, the timescale is calculated to be
\begin{eqnarray}
    t_\mathrm{diff}
    &\simeq&\left(\frac{3\kappa M_\mathrm{ej}}{4\pi cv_\mathrm{br}}\right)^{1/2}
    \nonumber\\
    &\simeq&
    100\ \mathrm{days}\left(\frac{\kappa}{0.2\ \mathrm{cm}^2\ \mathrm{g}^{-1}}\right)^{1/2}
    \nonumber\\&&\times
    \left(\frac{M_\mathrm{ej}}{10M_\odot}\right)^{1/2}
    \left(\frac{v_\mathrm{br}}{4.5\times 10^8\mathrm{cm}\ \mathrm{s}^{-1}}\right)^{-1/2},
    \label{eq:t_diff} 
\end{eqnarray}
\citep{1982ApJ...253..785A}, which we simply refer to as the diffusion timescale. 
We note that the increased ejecta velocity due to the central energy injection decreases the diffusion timescale. 

\subsection{Element transport}
We assume that initially the SN ejecta is approximately composed of four concentric layers with different chemical compositions rather than introducing the detailed distributions of specific elements. 
The four layers are referred to as the inner, intermediate 1, intermediate 2, and the outer layers. 
The freely expanding SN ejecta should have experienced steady and explosive nucleosynthesis before the initiation of the energy injection and thus each layer contains characteristic elements produced by steady and/or explosive nuclear burning episodes. 
Roughly speaking, the inner layer represents the innermost part abundant in iron-peak elements, such as $^{56}$Ni, the two intermediate layers represent regions with incomplete Si- and O-burning ashes, and the outer layer represent unburned stellar envelope dominated by carbon and oxygen. 

The mass fractions of the four layers are denoted by $X_\mathrm{in}$, $X_\mathrm{inter,1}$, $X_\mathrm{inter,2}$, and $X_\mathrm{out}$. 
The mass of the inner three layers are assumed to be $0.1$, $0.5$, and $0.5M_\odot$, respectively. 
Therefore we assume the following initial mass fraction distribution,
\begin{eqnarray}
    &&(X_\mathrm{in},X_\mathrm{inter,1},X_\mathrm{inter,2},X_\mathrm{out})
    \nonumber\\
    &&=
    \left\{\begin{array}{ccl}
    (1,0,0,0)&\ \ \mathrm{for}\ \ & M(R)/M_\odot\leq 0.1\\
    (0,1,0,0)&\ \ \mathrm{for}\ \ & 0.1<M(R)/M_\odot\leq 0.6\\
    (0,0,1,0)&\ \ \mathrm{for}\ \ & 0.6<M(R)/M_\odot\leq 1.1\\   
    (0,0,0,1)&\ \ \mathrm{for}\ \ & 1.1<M(R)/M_\odot\leq M_\mathrm{ej}\\  
    (0,0,0,0)&\ \ \mathrm{for}\ \ & M_\mathrm{ej}<M(R)/M_\odot\\ 
    \end{array}\right.,
    \label{eq:Xele}
\end{eqnarray}
where $M(R)$ is the enclosed mass within the radius $R$. 

\subsection{Energy injection}
We inject additional energy into the SN ejecta in the following way. 
As we have described in Section \ref{sec:radiative_transfer}, the additional energy is initially assumed to be in the form of non-thermal radiation. 
Except for the introduction of non-thermal radiation component, the procedure of the energy injection is mostly similar to our previous 3D hydrodynamic simulations.

First, we specify a spherical region at the center of the SN ejecta (referred to as the energy injection region). 
The outer radius of the energy injection region is fixed to a specific layer of the expanding SN ejecta and thus the region is expanding in the frame fixed at the center of the ejecta. 

Since the spin-down of a highly rotating magnetized neutron star is one of the most promising mechanisms powering bright SNSNe-I, we adopt the following functional form for the energy injection rate, 
\begin{equation}
    L_\mathrm{in}(t)=\frac{L_{\mathrm{in},0}}{(1+t/t_\mathrm{in})^2},
    \label{eq:Lin}
\end{equation}
which corresponds to the well-known dipole formula. 
Initially the energy is injected at an almost constant rate, $L_\mathrm{in}(t)\simeq L_\mathrm{in,0}$, at $t\ll t_\mathrm{in}$. 
Therefore, the timescale at which the injected energy reaches the initial kinetic energy, $t_\mathrm{c}=E_\mathrm{sn}/L_\mathrm{in,0}$ gives a convenient normalization time for the numerical simulations. 

We conduct simulations with different energy injection parameters, the initial injection rate $L_\mathrm{in,0}$ and the injection timescale $t_\mathrm{in}$. 
We particularly consider two model series with $L_\mathrm{in,0}=10^{46}$ and $10^{48}$ erg s$^{-1}$ ($t_\mathrm{c}=10^5$ and $10^3$ s, respectively). 
For each model series, we consider three models with different $t_\mathrm{in}/t_\mathrm{c}=1.0$, $3.0$, and $10.0$. 
The integration of Equation \ref{eq:Lin} gives the total injected energy $E_\mathrm{in}$,
\begin{equation}
    E_\mathrm{in}=L_\mathrm{in,0}t_\mathrm{in},
\end{equation}
at $t\gg t_\mathrm{in}$.
Therefore, the injected energies are $10^{51}$, $3\times 10^{51}$, and $10^{52}$ erg for the three models. 
Table \ref{table:model_description} summarizes the model names and parameters.

We also inject a small amount of gas along with the additional energy to avoid completely hollowing out the energy injection region and creating a central cavity with too small gas density, which is difficult to handle numerically. 
The mass injection rate $M_\mathrm{in}(t)$ is proportional to the energy injection rate;
\begin{equation}
    \dot{M}_\mathrm{in}(t)=\frac{L_\mathrm{in}(t)}{\eta c^2},
\end{equation}
with $\eta=20$. 
Then, the total injected mass $M_\mathrm{in}$ is given by
\begin{equation}
    M_\mathrm{in}\simeq 3\times10^{-4}M_\odot
    \left(\frac{E_\mathrm{in}}{10^{52}\ \mathrm{erg}}\right)
    \left(\frac{\eta}{20}\right)^{-1},
    \label{eq:Min}
\end{equation}
which is much smaller than the ejecta mass and therefore would have no significant impact on the dynamical evolution.

The simulation box for \verb|L46| models is $-6.4\times 10^{16}\ \mathrm{cm}\leq r,z\leq 6.4\times 10^{16}\ \mathrm{cm}$. 
These simulations follow the ejecta evolution up to $t=10^7$ s. 
A smaller simulation box of $-1.0\times 10^{16}\ \mathrm{cm}\leq r,z\leq 1.0\times 10^{16}\ \mathrm{cm}$ is adopted for \verb|L48| models. 
For the latter models, we stop the simulations at several $10^5$ s, when the fastest component of the ejecta reaches the numerical boundary.

In the framework of the dipole radiation of a magnetized neutron star, the injected energy $E_\mathrm{in}$ and the timescale $t_\mathrm{in}$ are expressed as a function of the initial frequency $\Omega_0$ (or the initial period $P_0=2\pi/\Omega_0$) and the dipole magnetic field strength $B_\mathrm{dip}$ in the following way;
\begin{eqnarray}
    E_\mathrm{in}&=&\frac{I_\mathrm{sn}\Omega_0^2}{2}
    \nonumber\\
    &\simeq&
    2\times10^{52}
    \left(\frac{I_\mathrm{ns}}{10^{45}\ \mathrm{g\ cm}^2}\right)
    \left(\frac{P_0}{1\ \mathrm{ms}}\right)^{-2}
    \ \mathrm{erg},    
\end{eqnarray}
and
\begin{eqnarray}
    t_\mathrm{in}&\simeq&\frac{6I_\mathrm{ns}c^3}{B^2R_\mathrm{ns}^6\Omega_0^2}
    \\
    &\simeq&4.1\times10^3
    \left(
    \frac{I_\mathrm{ns}}{10^{45}\ \mathrm{g\ cm}^2}
    \right)
    \left(
    \frac{B_\mathrm{dip}}{10^{15}\ \mathrm{G}}
    \right)^{-2}
    \\&&\hspace{4em}\times
    \left(
    \frac{R_\mathrm{ns}}{10\ \mathrm{km}}
    \right)^{-6}
    \left(
    \frac{P_0}{1\ \mathrm{ms}}
    \right)^2\ \mathrm{s},
    \nonumber
\end{eqnarray}
\citep[e.g.,][]{1983bhwd.book.....S}, where $I_\mathrm{ns}$ and $R_\mathrm{ns}$ are the moment of inertia and the radius of the neutron star. 
Therefore, the injected energy of $10^{51}$--$10^{52}$ erg corresponds to the initial spin period of the order of $1$ ms. 
The injection timescale of the order of $10^3$--$10^4$ s (\verb|L48| models) is realized by an extremely strong magnetic field strength, $B_\mathrm{dip}\sim 10^{15}$ G. 
Relatively weak magnetic fields of $B_\mathrm{dip}\sim 10^{14}$ G lead to longer injection timescales of $10^{5}$--$10^6$ s corresponding to \verb|L46| models. 
The parameters of \verb|L46| models are similar to the values inferred from the light curve fitting of SLSNe-I \citep[e.g.,][]{2017ApJ...850...55N}. 
The stronger magnetic field strength for \verb|L48| models, $B_\mathrm{dip}=10^{15}$ G or higher, would be achieved in magneto-rotational CCSNe \citep[e.g.,][]{2000ApJ...537..810W,2003ApJ...584..954A,2004ApJ...608..391K,2004ApJ...611..380T,2009ApJ...691.1360T}, which may be related to some broad-lined Type-Ic SNe. 

\begin{table}
\begin{center}
  \caption{Model descriptions}
\begin{tabular}{lrrr}
\hline\hline\\
Model name&$L_\mathrm{in,0}[\mathrm{erg}\ \mathrm{s}^{-1}]$&$t_\mathrm{in}[\mathrm{s}]$&$E_\mathrm{in}[10^{51}\mathrm{erg}]$
\\
\hline
\verb|L46small|&$10^{46}$&$1.0\times 10^5$&$1.0$\\
\verb|L46mid|&$10^{46}$&$3.0\times 10^5$&$3.0$\\
\verb|L46large|&$10^{46}$&$1.0\times 10^6$&$10.0$\\
\verb|L48small|&$10^{48}$&$1.0\times 10^3$&$1.0$\\
\verb|L48mid|&$10^{48}$&$3.0\times 10^3$&$3.0$\\
\verb|L48large|&$10^{48}$&$1.0\times 10^4$&$10.0$\\
\hline\hline
\end{tabular}
\label{table:model_description}
\end{center}
\end{table}

\section{Impact of a central power source\label{sec:results}}
In this section, we present the results of our numerical simulations. 
For the purpose of comparison, we also carry out 1D spherical simulations with the same initial setups as the \verb|L46| model series. 
We first present the results of 1D spherical simulations in Section \ref{sec:1d} and then those of 2D simulations in Sections \ref{sec:2d_L46} and \ref{sec:2d_L48}. 

We note that the central energy injection results in a forward shock propagating in the SN ejecta, while SN ejecta itself drives a shock in the ambient gas, which is usually called a forward shock. 
In the following, the shock driven by the central energy injection and propagating in the SN ejecta is simply referred to as the forward shock. 

\subsection{1D spherical models\label{sec:1d}}
\begin{figure*}
\begin{center}
\includegraphics[scale=0.55]{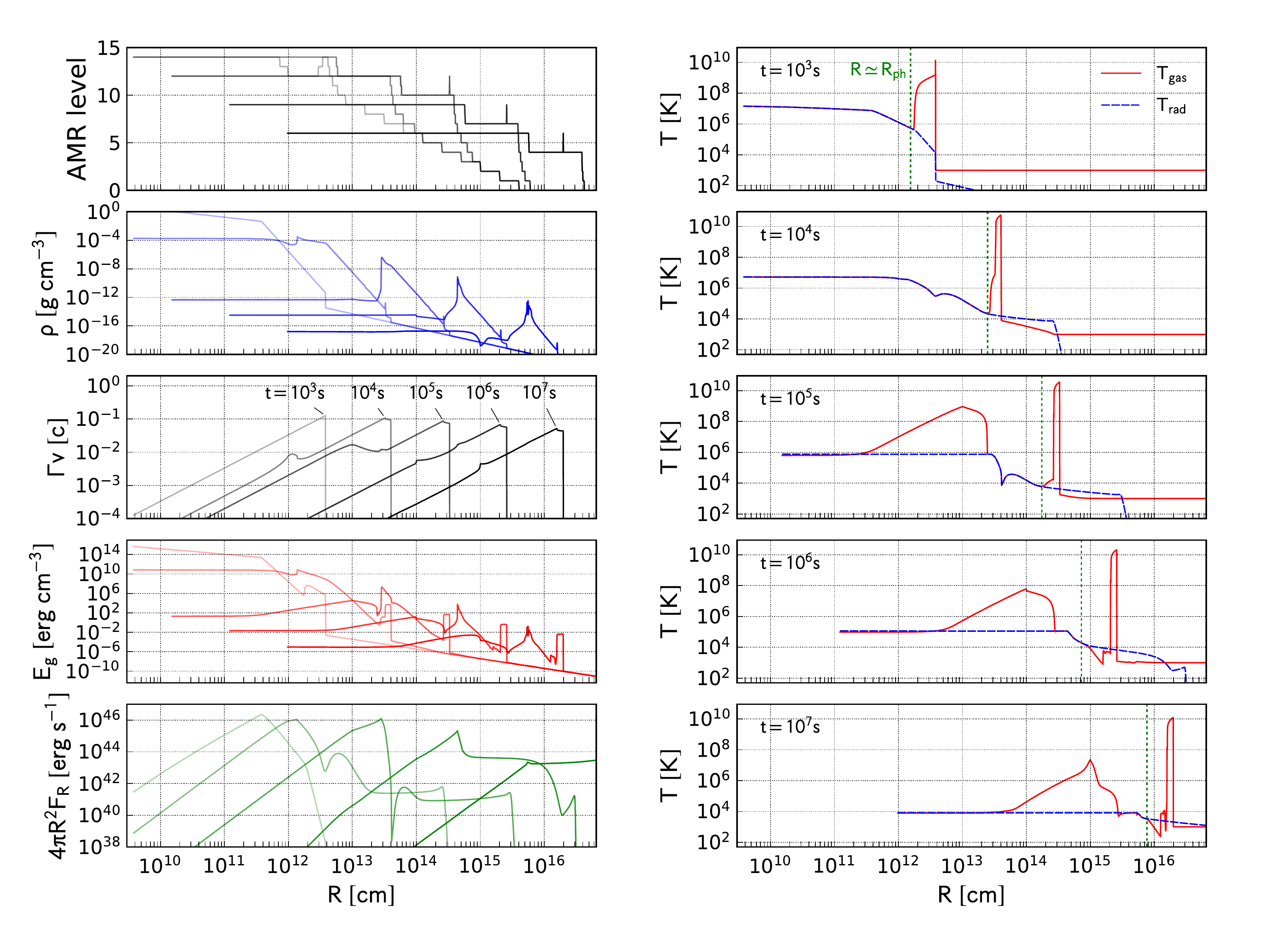}
\caption{Results of the 1D spherical simulation with $L_\mathrm{in,0}=10^{46}$ and $t_\mathrm{in}=3\times 10^5$ s. 
{\it Left:} the radial profiles of the AMR level, density, radial velocity, gas energy density, and the luminosity are plotted from top to bottom. 
The profiles at $t=10^3$, $10^4$, $10^5$, $10^6$, and $10^7$ s are shown in each panel. 
{\it Right: }
the gas (solid red line) and radiation (dashed blue line) temperature profiles at the same epochs are plotted from top to bottom. 
The vertical dotted line (green) in each panel represents the location of the photosphere, $R\simeq R_\mathrm{ph}$, above which gas and radiation are not effectively coupled. 
}
\label{fig:radial_1d}
\end{center}
\end{figure*}

\begin{figure*}
\begin{center}
\includegraphics[scale=0.55]{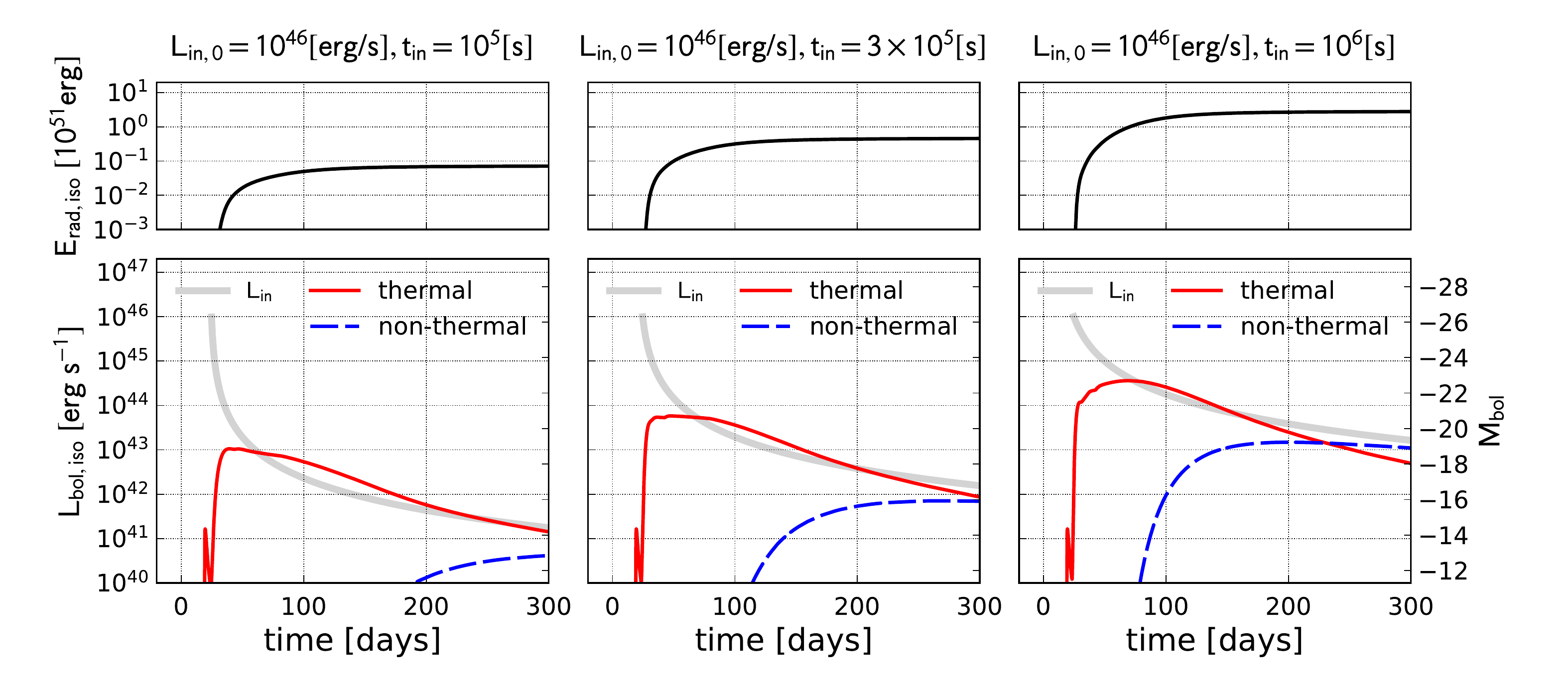}
\caption{Light curves of the spherical 1D models.
The lower panels show the bolometric light curves of the models with $t_\mathrm{in}=10^5$ (left), $3\times 10^5$ (center), and $10^6$ s (right). 
The light curves for the thermal and non-thermal components are plotted as solid and dashed lines. 
The energy injection rate is also plotted as a function of time. 
The upper panels show the corresponding cumulative radiated energy. }
\label{fig:lc_1d}
\end{center}
\end{figure*}

Figure \ref{fig:radial_1d} shows the evolution of the SN ejecta in the 1D spherical simulation with $L_\mathrm{in,0}=10^{46}$ erg s$^{-1}$ and $t_\mathrm{in}=3\times 10^5$ s (the 1D counterpart of model \verb|L46mid|). 
The dynamical evolution outlined below agrees with the previous 1D studies on SN ejecta with a central energy source \citep[e.g.,][]{2010ApJ...717..245K}. 
At $t=10^3$ s, i.e., right after the start of the simulation, the density profile simply follows the assumed double power-law function and the velocity is proportional to the radius. 
Then, the SN ejecta starts interacting with the injected and ambient gas at the innermost and outermost layers, respectively. 
The injected gas excavates the ejecta, producing a cavity filled with hot gas in the center (referred to as a ``hot bubble''). 
The inner ejecta pushed by the hot bubble is piling up to form a spherical shell, characterized by a sharp peak in the density profile at $t>10^4$ s, where most of the injected energy is converted to thermal radiation. 
Un-shocked SN ejecta stratified above the spherical shell is sufficiently dense to trap the thermal photons. 
As seen in the bottom panel of Figure \ref{fig:radial_1d}, the thermal radiation leaving the opaque ejecta shows a luminosity close to $10^{44}$ erg s$^{-1}$, which is high enough for explaining bright SLSNe-I. 

The importance of gas-radiation coupling is highlighted in the right panels of Figure \ref{fig:radial_1d}. 
The dense shell and the inner layers of the ejecta exhibit almost the same gas and radiation temperatures, indicating that the gas-radiation equilibrium is achieved. 
The inner part of the SN ejecta is heated by the injected energy and thus kept hot. 
At $t=10^6$ s, for example, the inner ejecta shows $T_\mathrm{g}\simeq 10^5$ K. 
At outer layers, however, the gas temperature significantly exceeds the radiation temperature. 
This is where the gas is only weakly coupled with radiation and therefore radiation pressure is not effective in pushing the layers ahead. 
The layer where the gas temperature starts deviating from the radiation temperature roughly corresponds to the photosphere separating the opaque and transparent parts of the ejecta (the vertical dotted lines in the right panels of Figure \ref{fig:radial_1d}). 
These features are reproduced only in radiation-hydrodynamic simulations where the coupling between gas and radiation is properly handled. 
At $t=10^6$ s, the temperature of the interface between the opaque and transparent parts is $T_\mathrm{g}\simeq T_\mathrm{r}\simeq 10^4$ K, which agrees with UV-dominated emission from SLSNe-I. 
The inner cavity also shows different gas and radiation temperatures. 
This is because the hot bubble blows the inner ejecta away and the density drops down to small values that cannot keep the hot bubble opaque.

We also performed spherical 1D simulations with the injection timescales of $t_\mathrm{in}=10^5$ and $10^6$ s. 
Their results are qualitatively similar to the model shown in Figure \ref{fig:radial_1d}. 
In Figure \ref{fig:lc_1d}, we plot the bolometric light curves for the three spherical models. 
An important trend is that the luminosity of the thermal emission initially dominates, while non-thermal luminosity becomes considerable at the late epochs. 
Since the energy injection is realized in the form of non-thermal radiation, this feature means that the injected energy is mostly converted to thermal radiation in the SN ejecta at early evolutionary stages. 
At later epoch, however, the conversion efficiency of the injected energy into thermal radiation drops down. 
Non-thermal radiation diffuses in the ejecta in a more extensive way as the ejecta density decreases and finally starts escaping into the surrounding space without interacting with the ejecta. 
This is how the non-thermal photon leakage happens in spherical symmetry. 

The model with the longest $t_\mathrm{in}$($=10^6\ \mathrm{s}$) shows the highest peak luminosity and the largest total radiated energy (the right panels of Figure \ref{fig:lc_1d}). 
This is due to the increased total injected energy. 
For the model with the shortest injection timescale ($t_\mathrm{in}=10^5$ s), the total injected energy, $E_\mathrm{in}=10^{51}$ erg, is only comparable to the initial kinetic energy of the SN ejecta. 
Therefore, the emission powered by the central energy injection is less luminous (the left panels of Figure \ref{fig:lc_1d}). 
The peak luminosities are $1.1\times10^{43}$, $5.8\times 10^{43}$, and $3.6\times 10^{44}$ erg s$^{-1}$ for the models with $t_\mathrm{in}=10^5$, $3\times 10^5$, and $10^6$ s, respectively. 
The corresponding total radiated energies are $7.1\times 10^{49}$, $4.6\times 10^{50}$, and $2.8\times 10^{51}$ erg. 
The leakage of non-thermal radiation also becomes more significant for larger injection energies. 
A larger injected energy efficiently accelerates the ejecta, making it dilute and transparent to non-thermal photons earlier. 
For the model with the largest injection energy (the right panel of Figure \ref{fig:lc_1d}), more than half of the injected energy leaves the ejecta without being converted to thermal radiation after $t\simeq 230$ days.  
This effect makes the thermal emission less luminous at later epoch and thus determines the declining rate of the late-time luminosity. 
We note that adopting a smaller non-thermal opacity $\kappa_\mathrm{nt}$ would enhance the non-thermal leakage and lead to a more drastic decline in the late-time light curve. 

\subsection{2D models with low energy injection rates\label{sec:2d_L46}}
In the following, we focus on the results of the 2D models with $L_\mathrm{in,0}=10^{46}$ erg s$^{-1}$.

\begin{figure*}
\begin{center}
\includegraphics[scale=0.10]{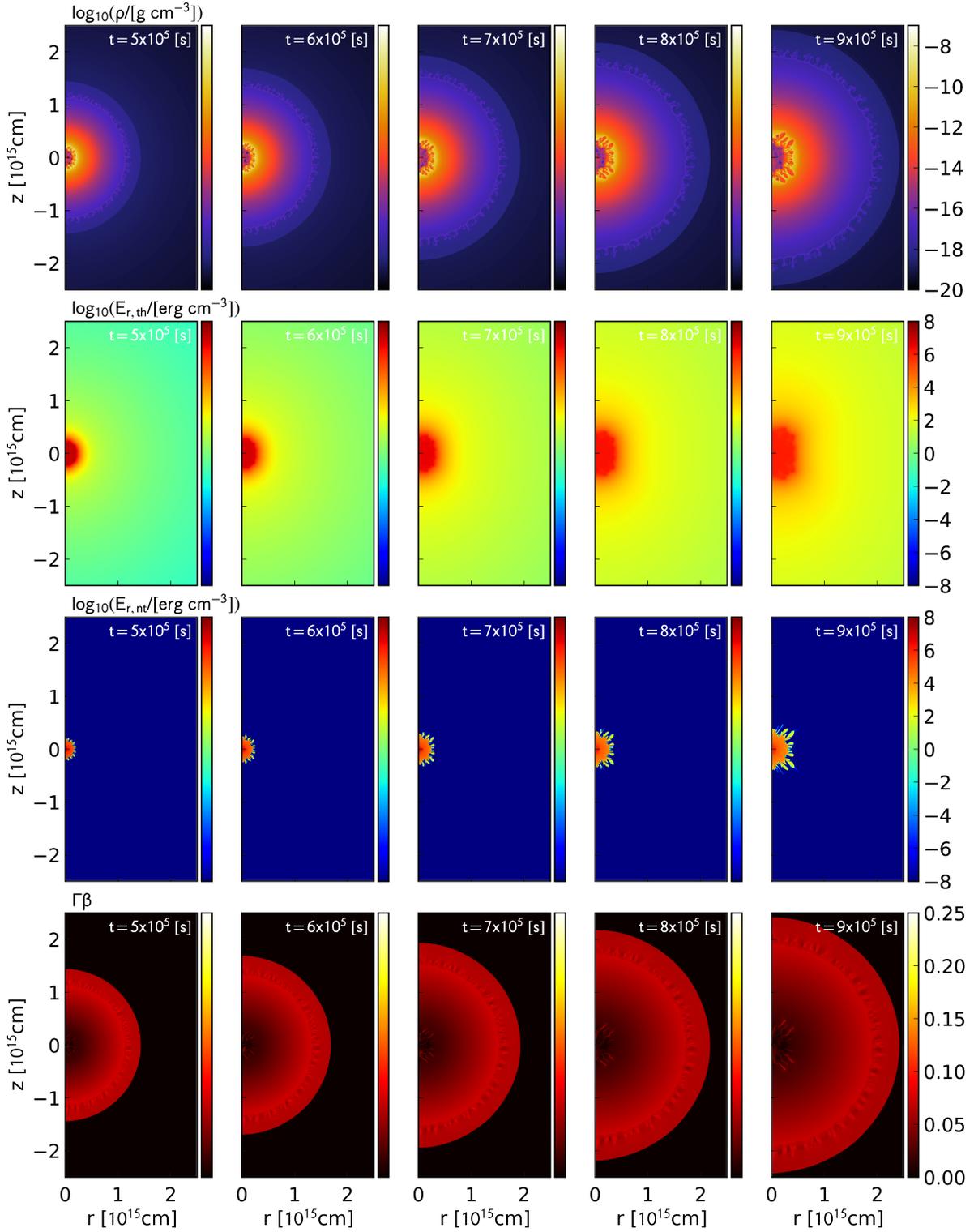}
\cprotect\caption{Results of model \verb|L46mid|. Spatial distributions of the density, thermal radiation energy density, non-thermal radiation energy density, and 4-velocity are plotted from top to bottom. 
The columns present distributions at $t=10^6$, $2\times10^6$, $3\times10^6$, $4\times10^6$, and $5\times10^6$ s from left to right. }
\label{fig:evol1_L46mid}
\end{center}
\end{figure*}
\begin{figure*}
\begin{center}
\includegraphics[scale=0.1]{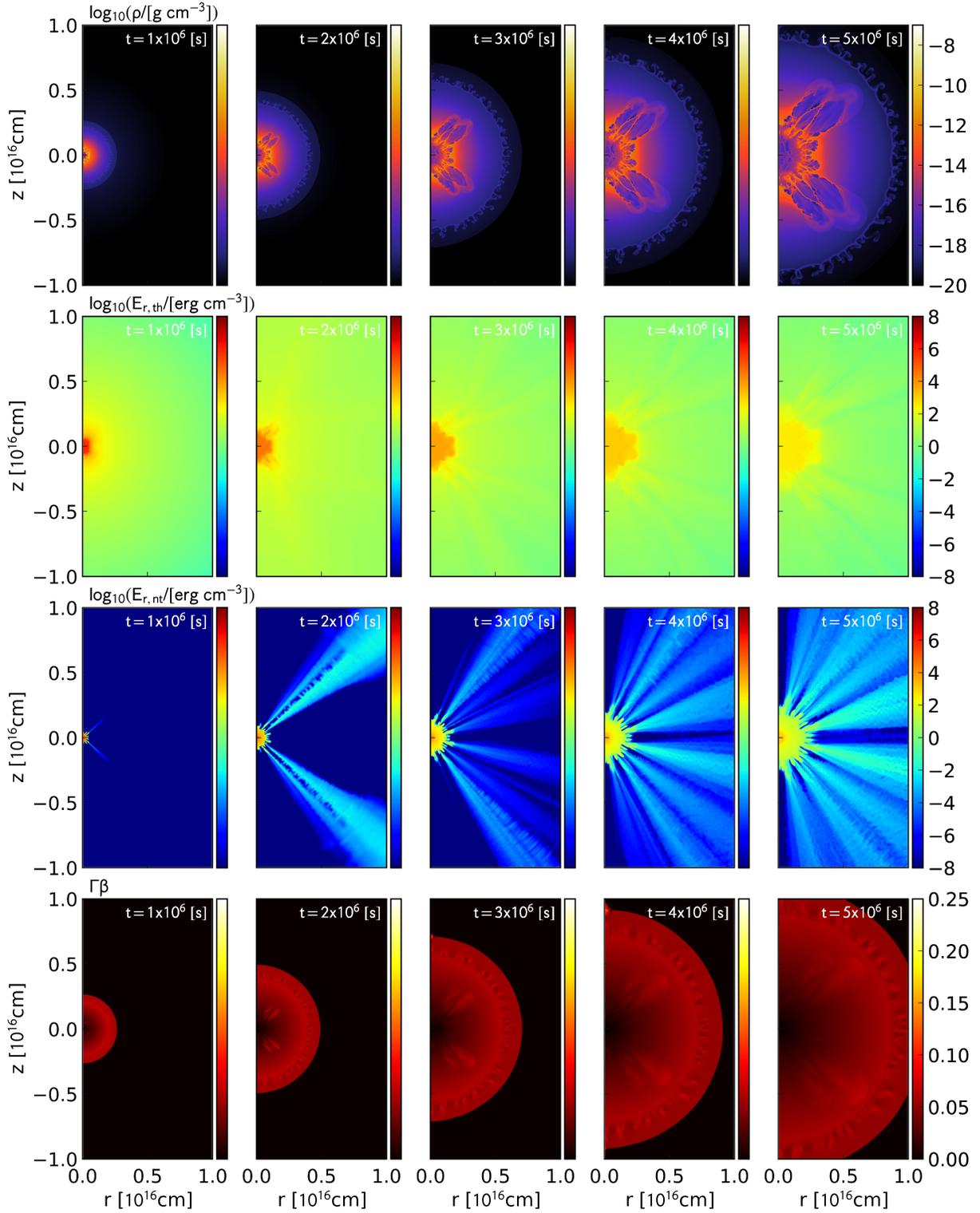}
\caption{Same as \ref{fig:evol1_L46mid}, but at later epochs.
The columns present distributions at $t=10^6$, $2\times10^6$, $3\times10^6$, $4\times10^6$, and $5\times10^6$ s from left to right.
}
\label{fig:evol2_L46mid}
\end{center}
\end{figure*}

\subsubsection{Dynamical evolution of ejecta and radiation field\label{sec:dynamical_evolution}}

Figures \ref{fig:evol1_L46mid} and \ref{fig:evol2_L46mid} show the dynamical evolution of the ejecta for model \verb|L46mid|. 
At early epochs, the SN ejecta is almost freely expanding into the surrounding space. 
Therefore, the ejecta initially keeps its spherical shape. 
As time elapses, however, the central energy injection and the associated expanding hot bubble gradually break the symmetry. 
The contact surface separating the ejecta and the injected gas is susceptible to the Rayleigh-Taylor instability, which amplifies perturbations at the interface. 
While the hot bubble is well confined in the inner region of the ejecta at $t<10^6$ s (Figure \ref{fig:evol1_L46mid}), the gas and radiation leaking from the hot bubble-ejecta interface push the outer layers away, leading to the destruction of the ejecta (Figure \ref{fig:evol2_L46mid}). 

The destruction of the SN ejecta has been recognized in the previous hydrodynamic simulations without radiative transfer effects \citep{2017MNRAS.466.2633S,2019ApJ...880..150S}. 
The hot bubble breakout in the radiation-hydrodynamic simulations happens in a similar way to the hydrodynamic simulations, because the inner ejecta is still optically thick. 
In other words, gas and radiation are tightly coupled in the inner region and the mixture behaves as a single fluid with an effective adiabatic index of $4/3$. 
The Rayleigh-Taylor instability grows and creates filamentary structures, which are often called Rayleigh-Taylor fingers. 
The outwardly propagating forward shock feels the ram pressure of the stratified layers. 
As long as the forward shock is within the inner ejecta, which is separated by the interface at $R=v_\mathrm{br}t$, the ram pressure is high enough to contain the hot bubble. 
Once the forward shock reaches the interface, however, the ram pressure of the outer ejecta is no longer high enough to contain the hot bubble because of the decreased density, resulting in the acceleration of the forward shock. 
The forward shock finally emerges from the photosphere in the ejecta. 

The SN ejecta after the hot bubble breakout shows filamentary structure. 
The outflows from the hot bubble penetrate the ejecta and their side-way expansion compresses the surrounding gas, making dense filaments. 
The Kelvin-Helmholtz instability could develop in the shear layers between the outflows and the ejecta, and also plays a role in further making the ejecta structure filamentary. 
In addition, denser gas is easier to cool by free-free emission because of its strong density dependence.  
Thus, the compressed gas is more likely to concentrate and form dense filaments in the late epochs. 
This is also an important effect of radiative transport that is not included in previous hydrodynamic simulations. 

One caveat on our simulation results is that the spatial distributions of hydrodynamic variables exhibit a global non-spherical structure elongated along the inclination angles of $45^\circ$ and $135^{\circ}$ even through the additional energy is injected in a spherical way. 
Our previous 2D simulation also showed a globally bipolar ejecta, which was artificially produced by the assumed axial symmetry \citep{2017MNRAS.466.2633S}. 
The different appearance of the ejecta in purely hydrodynamic and radiation-hydrodynamic simulations with 2D cylindrical geometry would be caused by the numerical treatment of gas and radiation. 
In the radiation-hydrodynamic simulations, it is the radiation pressure that pushes  the ejecta ahead of the hot bubble, and the transport of the radiation is carried out in a separate way from hydrodynamic variables. 
The radiation may prefer to penetrate in the inclination angles of $45^\circ$ and $135^\circ$ in the current treatment with M1 closure. 
In reality, as is investigated in our previous 3D hydrodynamic simulations \citep{2019ApJ...870...38S}, the shock structure is not clearly two-sided as is seen in Figures \ref{fig:evol1_L46mid} and \ref{fig:evol2_L46mid} but is composed of a number of outgoing flows from the hot bubble. 
This issue would be resolved in future 3D simulations with no assumed symmetry.  

The spatial distributions of both thermal and non-thermal radiation fields, which are not obtained in hydrodynamic simulations, are remarkably anisotropic particularly at late times.
This is also caused by the development of the Rayleigh-Taylor instability and the filamentary density structure. 
The growth of the Rayleigh-Taylor fingers produces low-density channels in the ejecta, through which both thermal and non-thermal photons can travel at a long distance without interacting with gas. 
As seen in the spatial distributions of the non-thermal radiation energy density in Figures \ref{fig:evol1_L46mid} and \ref{fig:evol2_L46mid}, non-thermal radiation propagates through the low-density channels like narrow pencil beams. 
This means that non-thermal radiation injected in the central region starts leaking from the ejecta after the hot bubble breakout. 
Since the non-thermal radiation is the energy source for powering the thermal radiation from the ejecta, this non-thermal radiation leakage decreases the conversion efficiency of the injected energy into thermal radiation and therefore significantly affects the bolometric light curve.

Thermal radiation is also affected by the ejecta structure. 
At the beginning of the hot bubble breakout (Figure \ref{fig:evol1_L46mid}), the outer space is filled with an almost uniform thermal radiation filed. 
This is because the thermal radiation is supplied by the hot bubble through the radiative diffusion in the ejecta. 
Although the radiation field in the hot bubble is anisotropic, the radiative diffusion makes it more isotropic as the radiation propagates throughout the ejecta. 
In the late epochs, however, the thermal radiation field becomes more anisotropic as in Figure \ref{fig:evol2_L46mid} according with the development of non-uniform density structure. 
Therefore, the thermal emission should exhibit significant viewing angle dependence at the late epochs. 

Finally, contrary to the previous hydrodynamic simulations, the hot bubble breakout is not effective in accelerating the outer layers of the ejecta. 
The spatial distributions of the 4-velocity in Figures \ref{fig:evol1_L46mid} and \ref{fig:evol2_L46mid} suggest that the velocities of the outermost layers are at most $\Gamma\beta\simeq 0.1$, although the previous hydrodynamic simulations showed a more efficient acceleration of the outermost layers up to $\Gamma\beta\simeq 1$ \citep{2017MNRAS.466.2633S,2019ApJ...870...38S}. 
In the hydrodynamic simulations, it is the forward shock that transports and deposits the injected energy in the outer ejecta. 
The forward shock is driven by the pressure of the hot bubble, which is dominated by radiation pressure. 
In fact, the radiative diffusion throughout the ejecta and the associated energy loss from the post-shock gas decrease the pressure of the hot bubble, which would have driven the fast forward shock. 
Furthermore, in the outermost layers, gas and radiation are already only weakly coupled at the moment of the hot bubble breakout. 
In other words, the photosphere is located in a halfway toward the outermost layer. 
Once the forward shock emerges from the photosphere, the shock releases the radiation energy stored in the post-shock layer and no longer is capable of accelerating the stratified layers to relativistic velocities. 
We shall further discuss this effect in comparison with the models with the high energy injection rate of $L_\mathrm{in,0}=10^{48}$ erg s$^{-1}$ in Section \ref{sec:2d_L48}. 

\subsubsection{Dependence on the injected energy}
\begin{figure}
\begin{center}
\includegraphics[scale=0.26]{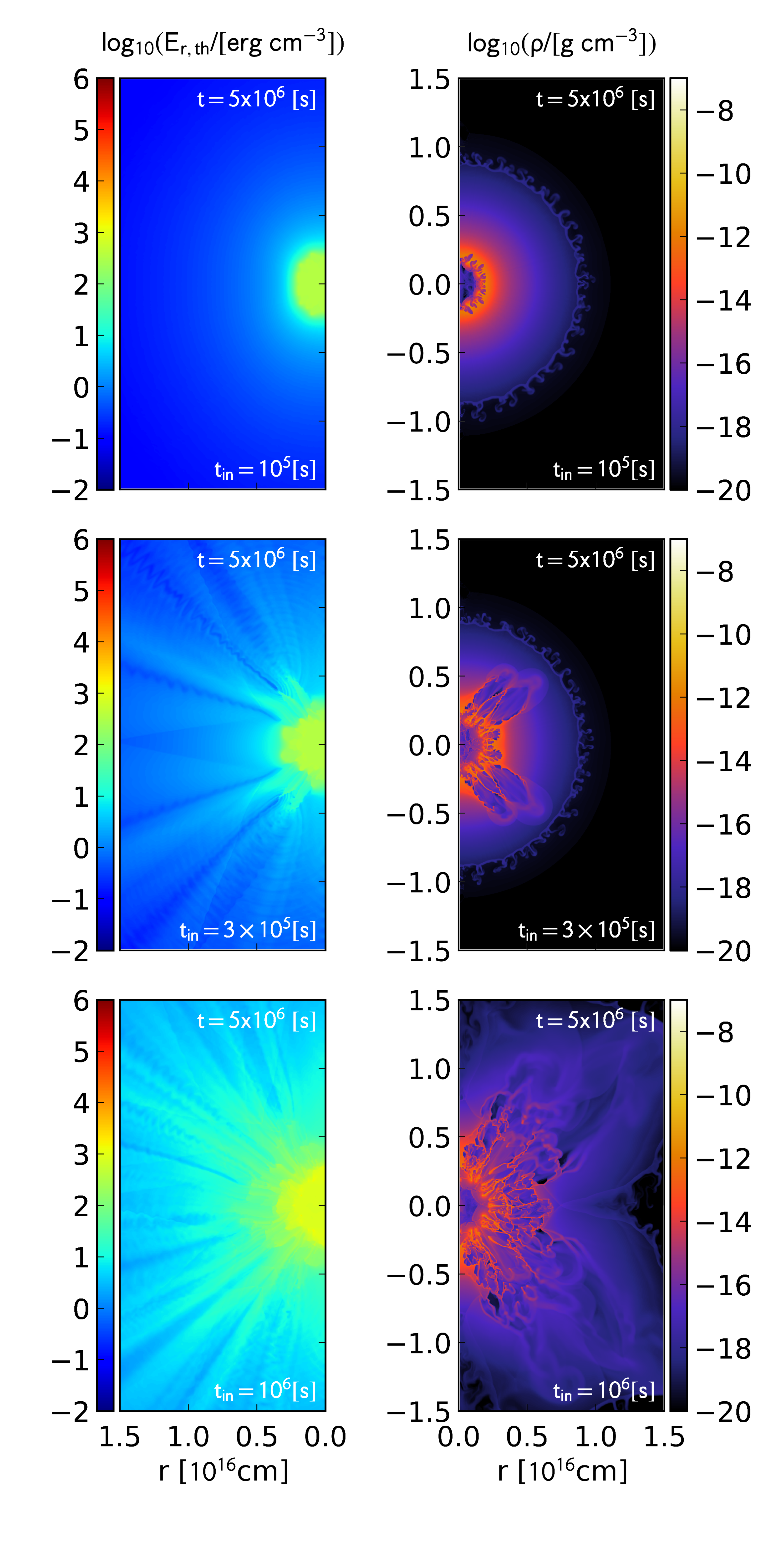}
\cprotect\caption{Comparison of the models with $L_\mathrm{in,0}=10^{46}$ erg s$^{-1}$. 
The mass density (right) and the thermal radiation energy (left) distributions at $t=5\times 10^6$ s are plotted for models \verb|L46small|, \verb|L46mid|, and \verb|L46large| from top to bottom. 
}
\label{fig:comparison_L46}
\end{center}
\end{figure}

The effect of the energy injection on the density structure is sensitive to the amount of the injected energy. 
In Figure \ref{fig:comparison_L46}, we compare the ejecta density structure and the radiation fields for the models with different $t_\mathrm{in}$ long after the injection time. 
As seen in the top panels of Figure \ref{fig:comparison_L46}, in model \verb|L46small|, a smaller part of the SN ejecta experiences the violent mixing compared with model \verb|L46mid|. 
The hot bubble expansion is not so intense to penetrate the whole ejecta in this least energetic model, leaving the outer part of the ejecta well-stratified. 
The limited effect of the hot bubble expansion with an insufficient energy has also been investigated in the 3D hydrodynamic simulation \citep{2019ApJ...880..150S}. 
The clearly separated inner mixing region and the outer stratified layers may be revealed by the spectral evolution. 
As seen in the left panels of Figure \ref{fig:comparison_L46}, the thermal radiation field outside the mixing layer is less anisotropic than the more energetic counterparts. 
We therefore expect a less prominent viewing angle effect in the bolometric light curves in this model, which is indeed confirmed below (see Section \ref{sec:lc}). 

In the most energetic model \verb|L46large| (the  bottom panels), on the other hand, the hot bubble breakout blows away the ejecta in a more violent way, tearing the ejecta apart into tiny pieces. In this model, the chemical elements deeply embedded in the ejecta, such as iron and nickel, could be mixed into the outer layers more efficiently, which should have a significant impact on the spectral evolution. 

\begin{figure*}
\begin{center}
\includegraphics[scale=0.5]{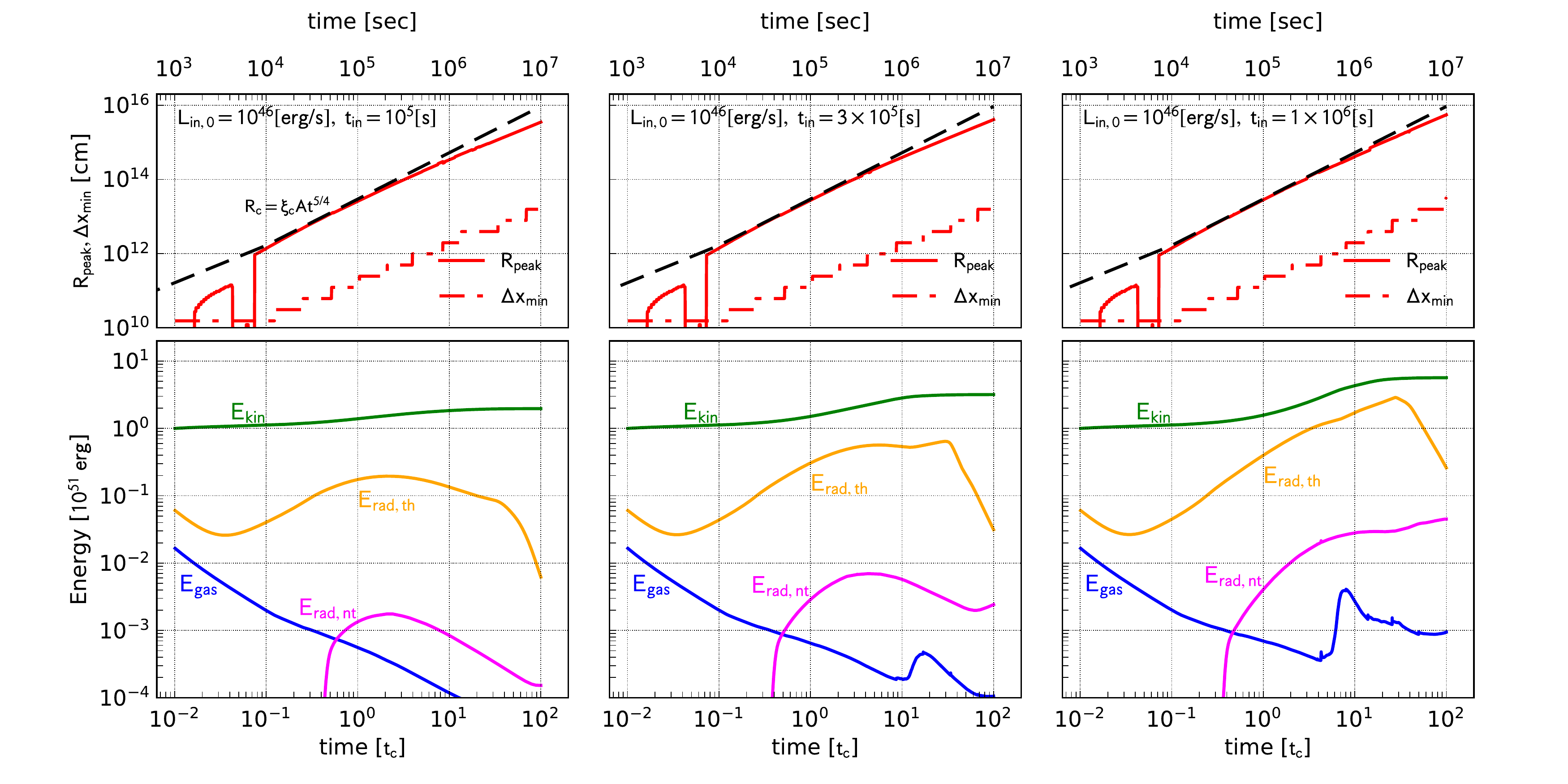}
\cprotect\caption{Temporal evolution of physical variables. 
In the upper panels, the radius $R_\mathrm{peak}$ at the density peak and the minimum resolved length $\Delta x_\mathrm{min}$ are plotted as a function of time (in units of $t_\mathrm{c}$). 
The temporal increase of $R_\mathrm{peak}$ is compared with a theoretical expectation $R_\mathrm{c}$ based on a self-similar solution. 
In the lower panels, the kinetic, the internal, and the radiation energies are plotted as a function of time. 
For radiation, we show the energies of thermal and non-thermal radiation components separately. 
The models \verb|L46small|, \verb|L46mid|, and \verb|L46large| are presented from left to right. }
\label{fig:energy_L46}
\end{center}
\end{figure*}

Figure \ref{fig:energy_L46} shows the temporal evolution of the characteristic radius and the energies in different forms; kinetic, thermal radiation, non-thermal radiation, and the gas internal energies. 
As is done in our previous work \citep{2017MNRAS.466.2633S,2019ApJ...880..150S}, we identify the numerical cell with the highest density in the numerical domain, which is referred to as the density peak, and keep track of the radius of the density peak. 
The density peak basically follows the position of the thin shell swept up by the expanding hot bubble. 
The dynamical evolution of the thin shell with a constant energy injection is well described by a series of self-similar solutions \citep{1992ApJ...395..540C,1998ApJ...499..282J}. 
For the parameter set adopted here, the radius is expressed as a power-law function of the elapsed time with an exponent of $5/4$. 
We also plot the time dependence expected from the self-similar solution in the upper panel of Figure \ref{fig:energy_L46}. 
The temporal evolution of the density peak shows a good agreement with the self-similar expansion law as long as the almost constant energy injection continues and the thin shell is quasi-spherical. 
In the models shown in Figure \ref{fig:energy_L46}, the radius of the density peak starts deviating from the power-law evolution after the injection timescale $t=t_\mathrm{in}$.

The kinetic energy is always dominating the other forms of energy. 
It increases according to the energy injection and then approaches the terminal value of $\simeq E_\mathrm{sn}+E_\mathrm{in}$ after $t=t_\mathrm{in}$. 
The gas internal energy is always overwhelmed by the thermal radiation energy and thus only plays a minor role in the dynamical evolution of the system. 
In other words, it is the radiation pressure of the shocked gas that pushes the ejecta. 
The thermal radiation energy first decreases according to the expansion of the ejecta at $t<0.1t_\mathrm{c}$.
This is because the radiation is initially strongly coupled with the ejecta and suffers from adiabatic cooling. 
Then, the thermal radiation energy starts increasing as the energy injection continues. 
The sudden drop in the thermal radiation energy in the final stage of its evolution ($t>3t_\mathrm{c}$) is caused by the escape of thermal radiation from the numerical domain. 
On the other hand, the amount of non-thermal radiation energy is initially negligible. 
The central part of the ejecta is dense enough to absorb almost all the non-thermal radiation immediately after the injection. 
It starts growing rapidly at around $t\sim 0.3t_\mathrm{c}$ as the ejecta becomes transparent to non-thermal photons.

\subsubsection{Material mixing\label{sec:mixing}}
\begin{figure*}
\begin{center}
\includegraphics[scale=0.4]{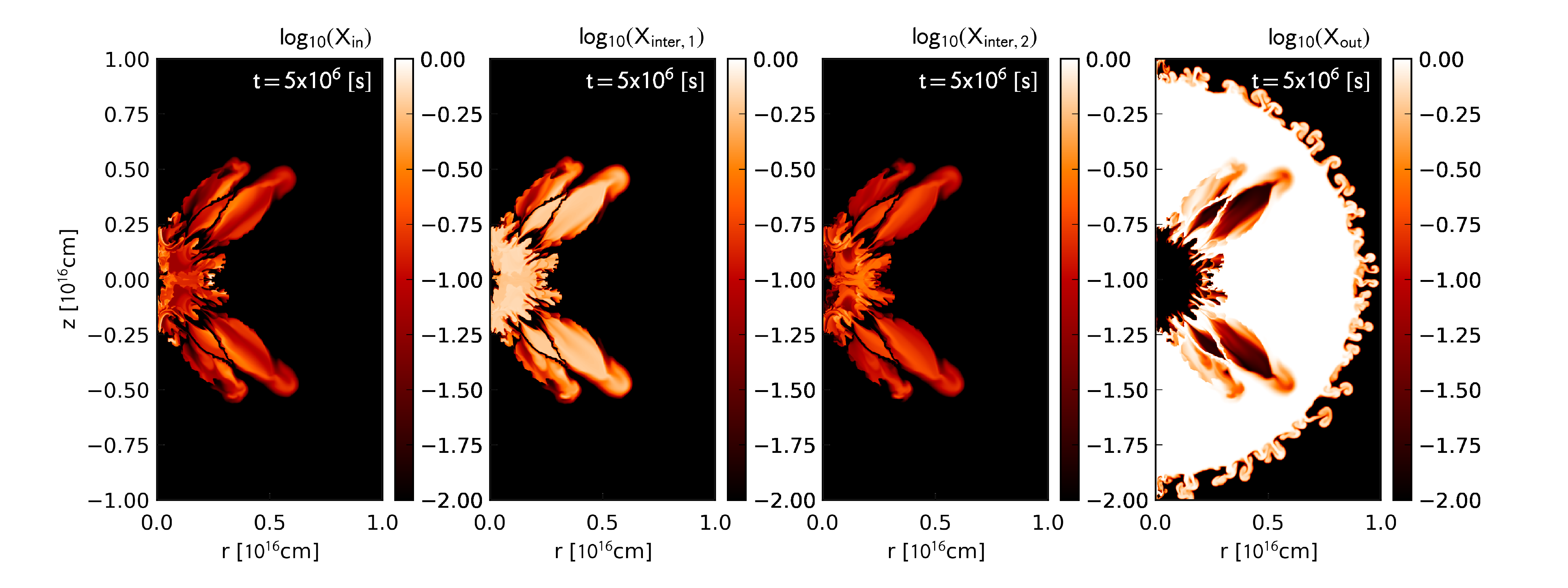}
\cprotect\caption{Material mixing caused by the central energy injection. 
From left to right, the panels represent the spatial distributions of the mass fraction, $X_\mathrm{in}$, $X_\mathrm{inter,1}$, $X_\mathrm{inter,2}$, and $X_\mathrm{out}$, defined in Equation \ref{eq:Xele} at $t=5\times 10^6$ s for model \verb|L46mid|. }
\label{fig:mixing}
\end{center}
\end{figure*}

The central energy injection and the ejecta destruction result in the efficient mixing of materials throughout the SN ejecta. 
As we defined in Equation \ref{eq:Xele}, we consider concentric layers with different chemical compositions in the ejecta. 
Without the central energy injection, these layers would expand freely and remain well-stratified. 
In the presence of the central engine, however, the hot bubble breakout makes these layers well mixed. 
Figure \ref{fig:mixing} shows how these different layers are distributed after the impact of the central energy injection for model \verb|L46mid|. 
The fast gas flow from the central energy source transports materials in the inner embedded layers toward the outer part of the ejecta, realizing the ``chemical inversion''. 
As the similar distributions of the inner and the two intermediate layers in Figure \ref{fig:mixing} suggest, the inner $\sim 1M_\odot$ of the ejecta is almost uniformly mixed. 
The distribution of these three layers basically follows the structure created by the development of the Rayleigh-Taylor fingers. 
We have seen that the Rayleigh-Taylor fingers most strongly develop along the inclination angles of $45^\circ$ and $135^\circ$. 
Although the density structure is highly sensitive to the configuration of the numerical simulations and admittedly suffers from the artificial symmetry, we can safely expect that chemical elements having been present in the innermost layer would be mixed into low-density regions created by the penetration of the outflows. 
Therefore, this chemical inversion makes it possible for such elements to be observed even in early evolutionary stages of SNe having experienced the hot bubble breakout. 
This finding has an impact on the spectral evolution of engine-driven SNe, which will be discussed in more detail in Section \ref{sec:discussion}.

\begin{figure}
\begin{center}
\includegraphics[scale=0.6]{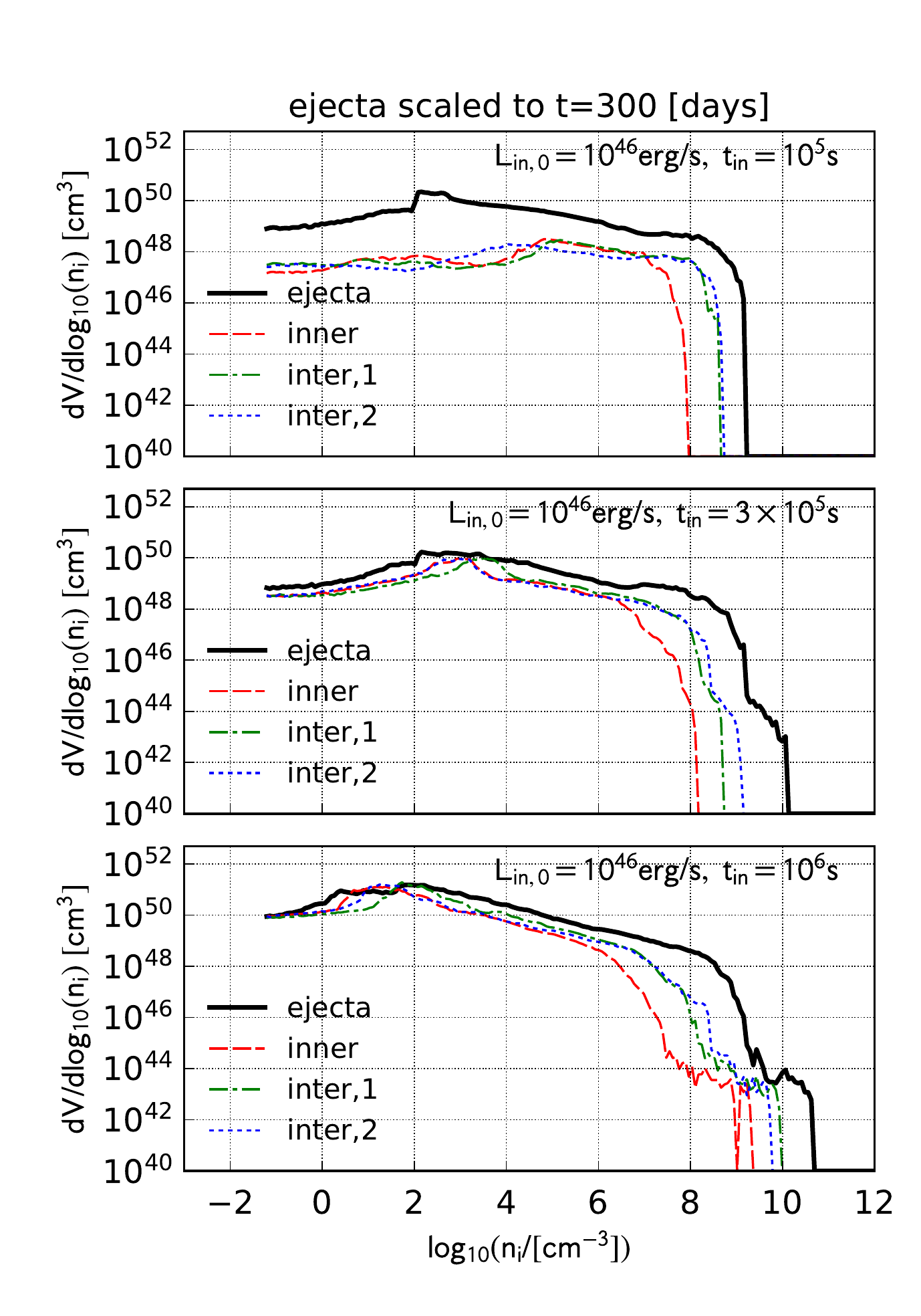}
\cprotect\caption{Differential volume as a function of the ion number density. 
The volume taken up by ejecta with given ion number densities are plotted for models \verb|L46small| (top), \verb|L46mid| (middle), and \verb|L46large| (bottom). 
The thick black line represents the relation for the whole ejecta, while thin dashed, dash-dotted, and dotted lines represent those for the inner and the two intermediate layers. }
\label{fig:dVdlog_n}
\end{center}
\end{figure}

A more quantitative look at the ejecta structure and the material mixing is obtained by the differential volume taken up by the ejecta with a given ion number density. 
The local ion number density at a numerical cell $(r_i,z_j)$ with a mass density $\rho_{ij}$ is defined as $n_\mathrm{i}(r_i,z_j)=\rho_{ij}/(\bar{A}m_\mathrm{u})$, by assuming fully ionized gas with an average mass number of $\bar{A}$. 
Then, for a given value of the ion number density $n_\mathrm{i}$, we calculate the following quantity by summing up the volume of numerical cells with their ion number densities in a certain logarithmic range $[\log_{10}n_\mathrm{i}-\Delta \log_{10}n_\mathrm{i}/2,n_\mathrm{i}+\Delta \log_{10}n_\mathrm{i}/2]$,
\begin{eqnarray}
    \frac{dV}{d\log_{10}n_\mathrm{i}}&=&
    \sum_{ij} 2\pi r_i\Delta r_i\Delta  z_j
    \\&&
    \times{\cal W}(\log_{10}[\rho_{ij}/(\bar{A}m_\mathrm{u})],\log_{10}n_\mathrm{i},\Delta \log_{10}n_\mathrm{i})
    \nonumber
\end{eqnarray}
where ${\cal W}(x,y,\Delta y)$ is a window function defined as
\begin{equation}
    {\cal W}(x,y,\Delta y)=
    \left\{
    \begin{array}{cl}
    1&\mathrm{for}\ \ y-\Delta y/2\leq x\leq y+\Delta y/2,\\
    0&\mathrm{otherwise}.
    \end{array}
    \right.
\end{equation}
This differential volume is calculated by using the density distributions of the inner layer $\rho X_\mathrm{in}$, the two intermediate layers, $\rho X_\mathrm{inter,1}$ and $\rho X_\mathrm{inter,2}$, and the whole ejecta at $t=5\times 10^6$ s. 
Since the energy injection is no longer effective at $t=5\times 10^6$ s, we simply scale the density distribution at the epoch to $t=300$ days, at which nebular spectra are available for some well-observed SLSNe-I, by assuming free expansion. 

The distributions for models \verb|L46small|, \verb|L46mid|, and \verb|L46large| are plotted in Figure \ref{fig:dVdlog_n}. 
The largest volume is occupied by low-density gas with the ion number density of $n_\mathrm{i}=10^1$--$10^3$ cm$^{-3}$, where the differential volume distributions in Figure \ref{fig:dVdlog_n} show a peak. 
The distributions extend to the highest ion number density of $n_\mathrm{i}=10^8$--$10^9$ cm$^{-3}$, at which the distributions exhibit a break. 
The gas at the highest ion number density corresponds to the inner dense filaments found in the density distribution (top panels of Figure \ref{fig:evol2_L46mid}). 
Around the break, $n_\mathrm{i}=10^8$--$10^9$ cm$^{-3}$, the differential volume is smaller than the peak value by a factor of more than 100. 
This indicates that the inner dense filaments take up a much smaller volume than the rest of the ejecta and their filling factor is less than 0.01. 
The presence of such high-density filaments with a small filling factor is also important for the spectral formation of engine-driven SNe particularly at the nebular phase. 

The inner ejecta mixed into outer layers is selectively located in low-density channels created by the hot bubble breakout. 
In general, the differential volume distributions for the inner and intermediate layers in Figure \ref{fig:dVdlog_n} certainly do not extend up to the highest ion number density of $n_\mathrm{i}=10^8$--$10^9$ cm$^{-3}$. 
This trend is particularly true for the inner layer (dashed red lines in the panels of Figure \ref{fig:dVdlog_n}). 
Therefore, the composition of high-density filaments is dominated by unburned materials (e.g., oxygen and carbon) in outer layers, which is also clear from the spatial abundance distribution of the outer layer in Figure \ref{fig:mixing} (right panel). 
A clear difference among the models shown in Figure \ref{fig:dVdlog_n} is the volume taken up by the inner and the intermediate layers around the peak. 
For models \verb|L46mid| and \verb|L46large|, the distributions for the inner and the intermediate layers show their peaks at a similar density, $n_\mathrm{i}=10^1$--$10^3$ cm$^{-3}$, to those for the whole ejecta. 
In other words, both inner and intermediate layers are mixed into the outer layers with low densities. 
For model \verb|L46small|, in contrast, low-density materials do not take up a large volume at $n_\mathrm{i}=10^1$--$10^3$ cm$^{-3}$ and the peak is located around $n_\mathrm{i}=10^5$ cm$^{-3}$. 
This difference reflects the different extent of the mixing caused by the central energy injection. 
In model \verb|L46small|, only the inner part of the ejecta is stirred by the energy injection. 
Therefore, materials initially in the inner layers remain in the inner part of the ejecta.

Finally, we note the possible effect of mass injection. 
In our simulations, the energy injection is accompanied by a small amount of gas with a mass up to $\simeq 3\times 10^{-4}\ M_\odot$ (Equation \ref{eq:Min}). 
Since the injected mass is much smaller than that of the inner layer, $0.1\ M_\odot$, its dynamical impact is negligible. 
We expect that such a small amount of injected gas, if any, is distributed in the low-density region excavated by the hot bubble and the Rayleigh-Taylor fingers, resulting in the mass fraction distribution similar to that of the inner layer (the left panel of Figure \ref{fig:mixing}). 
Given that the central energy source would be a stellar-mass compact remnant, it is unlikely that the energy injection is accompanied by gas with a comparable mass to the SN ejecta. 
Nevertheless, gas injection with a similar mass to the inner layer, $\sim0.1\ M_\odot$, may be expected in the form of a wind from a newborn neutron star or a disk wind from a black hole-accretion disk system. 
When the injected material has a characteristic chemical composition different from the inner layers of the SN ejecta, it may significantly affect the resultant chemical structure in the ejecta.

\subsubsection{Bolometric light curves\label{sec:lc}}
\begin{figure*}[htbp]
\begin{center}
\includegraphics[scale=0.5]{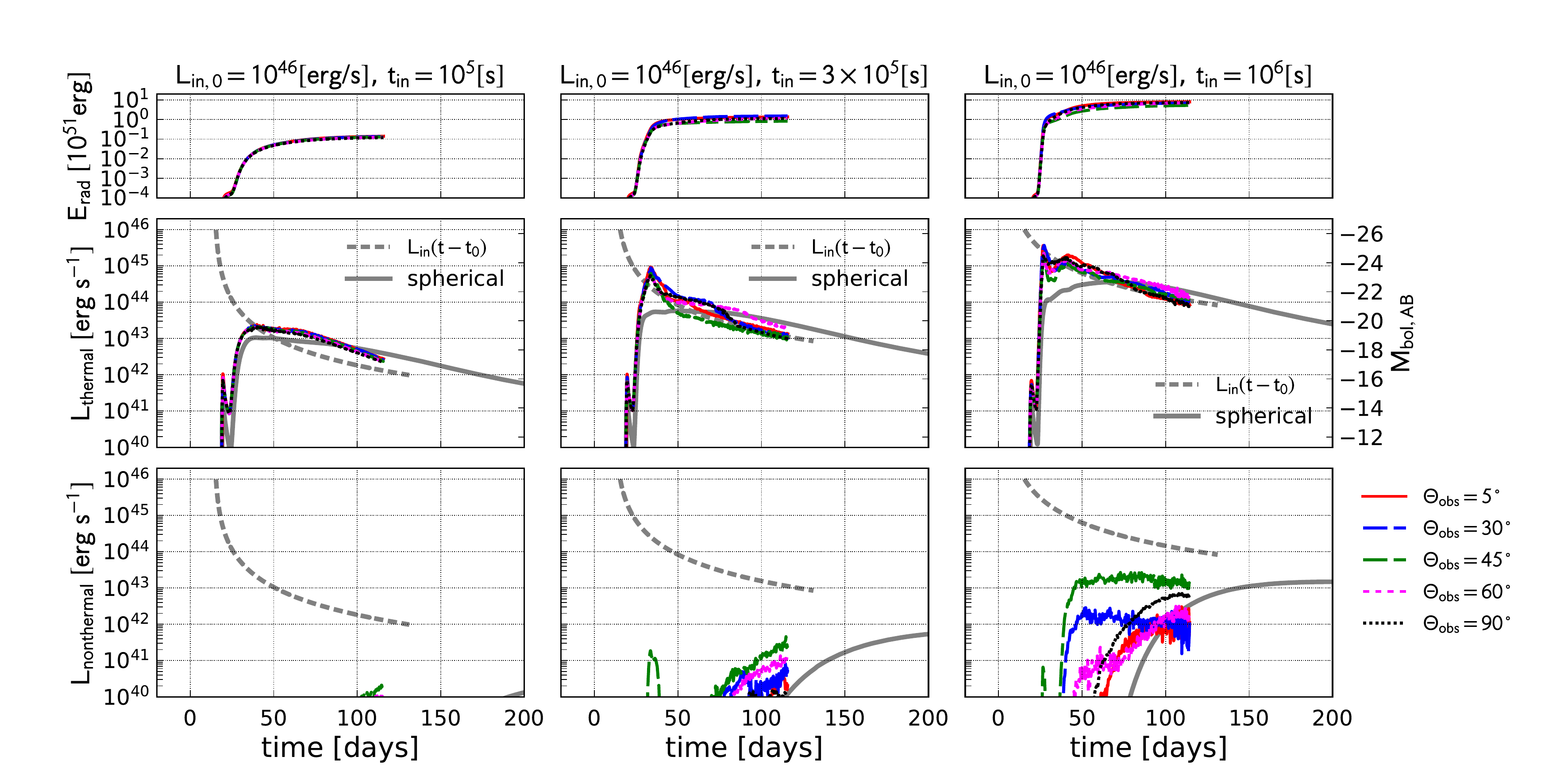}
\cprotect\caption{Light curves of models \verb|L46small| (left), \verb|L46mid| (middle), and \verb|L46large| (right). 
In each column, the cumulative radiated energy, the bolometric light curves for thermal and non-thermal radiation components are plotted from top to bottom. 
The light curves with the viewing angles of $\Theta_\mathrm{v}=5^\circ$, $30^\circ$, $45^\circ$, $60^\circ$, and $90^\circ$ are plotted. 
The energy injection rate (thick dashed) and the thermal and non-thermal light curves of the corresponding spherical models (thick) are also plotted for the purpose of comparison. }
\label{fig:lc_L46}
\end{center}
\end{figure*}

Figure \ref{fig:lc_L46} shows the bolometric light curves of the three models with $L_\mathrm{in,0}=10^{46}$ erg s$^{-1}$. 
The thermal emission can be much brighter than the peak luminosity of normal SNe, $\sim 10^{43}$ erg s$^{-1}$, as in the 1D spherical models. 
However, some qualitative differences are found between the 1D spherical and 2D models. 
First of all, the 2D models are more luminous and rapidly evolving than the 1D counterparts. 
This behavior reflects the qualitatively different dynamical evolution of the SN ejecta in 1D and 2D configurations. 
We focus on the two models with the higher energy injection rate, \verb|L46mid| and \verb|L46large|, both of which show the violent hot bubble breakout.  
As we have seen in Figures \ref{fig:evol1_L46mid} and \ref{fig:evol2_L46mid}, the destruction of the ejecta due to the hot bubble breakout makes it easier for photons to diffuse in the ejecta. 
The hot bubble breakout itself produces an intense flash of photons having been trapped inside the hot bubble in analogy to the shock breakout in normal SNe. 
Thus, the hot bubble breakout creates a sharp peak with a luminosity as bright as $L_\mathrm{bol}\simeq 10^{45}$ erg s$^{-1}$ at the beginning of the light curve evolution as seen in Figure \ref{fig:lc_L46} (right coulum, in particular). 
This shock-powered initial emission and the resultant double-peaked light curve have already been pointed by the 1D spherical simulations by \cite{2016ApJ...821...36K}, who called this phenomenon the ``magnetar-driven shock breakout''. 
As demonstrated by \cite{2016ApJ...821...36K}, the breakout component is sometimes dim and therefore is not easily separable from the main component depending on the adopted model parameters, such as the ejecta mass and energy, as well as the thermalization efficiency of the injected energy. 
In our 1D simulations shown in Figure \ref{fig:lc_1d}, the breakout component is not prominent compared to the main peak.  
In a multi-dimensional setting with large injected energies, however, the breakout happens in a much more drastic way, making the associated emission bright.

After the initial bright peak, a moderately decaying part or a second bump appears. 
The decay rate of the light curve in this part is broadly consistent with the energy injection rate and therefore is powered by the continuous energy injection at the center.  
Even aside from the first peak, which is not prominent in the 1D counterpart, this part following the first peak evolves more rapidly in the 2D models compared to the  corresponding 1D light curves. 
In the 1D spherical models, the expanding hot bubble simply piles up the inner layers of the ejecta and creates a geometrically thin shell (see Figure \ref{fig:radial_1d}), which efficiently traps thermal and non-thermal photons. 
Therefore, it takes a long time for the injected energy to diffuse through the ejecta. 
This results in the slowly evolving  1D light curves with the evolutionary timescales of several 10 days. 
The rapid light curve evolution in 2D models indicates that the timescales of the energy transport in the ejecta is considerably short. 
Non-thermal photons can easily travel through the holes created by the hot bubble breakout and directly heat the outer layer of the ejecta. 
In other words, the filamentary density structure increases the effective photon mean free path and shortens the diffusion timescale. 
This effect is also found in the non-thermal light curves, which exhibit earlier rises compared with 1D counterparts. 
As expected from the spatial distribution of non-thermal radiation energy density in Figures \ref{fig:evol1_L46mid} and \ref{fig:evol2_L46mid}, the leakage of non-thermal radiation from the ejecta happens much earlier in the 2D models than the 1D counterparts. 

The filamentary density structure may also cause the high time variability and the strong viewing angle dependence of the light curves. 
As the density of the ejecta is reduced along some specific directions due to the outflow penetration, the photospheric radius along different radial directions are highly angular dependent. 
While an observer can see the deep interior of the ejecta through a low-density channel created by the hot bubble breakout, the inner region of the ejecta may be hidden for another observer with a different line of sight. 
The difference in the light curves with different viewing angles is more significant for the models with higher energy injection rates, \verb|L46mid| and \verb|L46large|, which reflects the large amount of the injected energy and its impact on the density structure. 
The viewing angle dependence of the light curves for model \verb|L46small| is less prominent. 
As seen in the top panels of Figure \ref{fig:comparison_L46}, for \verb|L46small|, only the inner ejecta have been affected by the Rayleigh-Taylor mixing even at several 10 days. 
The photosphere is located in the unaffected outer layer of the ejecta, while the mixing region is still deeply embedded. 
In this case, photons produced in the hot bubble first experience filamentary density structure on their way out of the ejecta, where their angular distribution could be anisotropic. 
However, the inner anisotropy would be erased and thus the viewing angle difference would be smeared out while those photons diffuse through the well-stratified outer layers with multiple scattering and absorption episodes. 

\subsection{2D models with high energy injection rates\label{sec:2d_L48}}
In this subsection, we present the simulation results of the models with the higher energy injection rate $L_\mathrm{in,0}=10^{48}$ erg s$^{-1}$ with a particular focus on differences from the models with $L_\mathrm{in,0}=10^{46}$ erg s$^{-1}$ presented in the previous subsection. 

\begin{figure*}
\begin{center}
\includegraphics[scale=0.4]{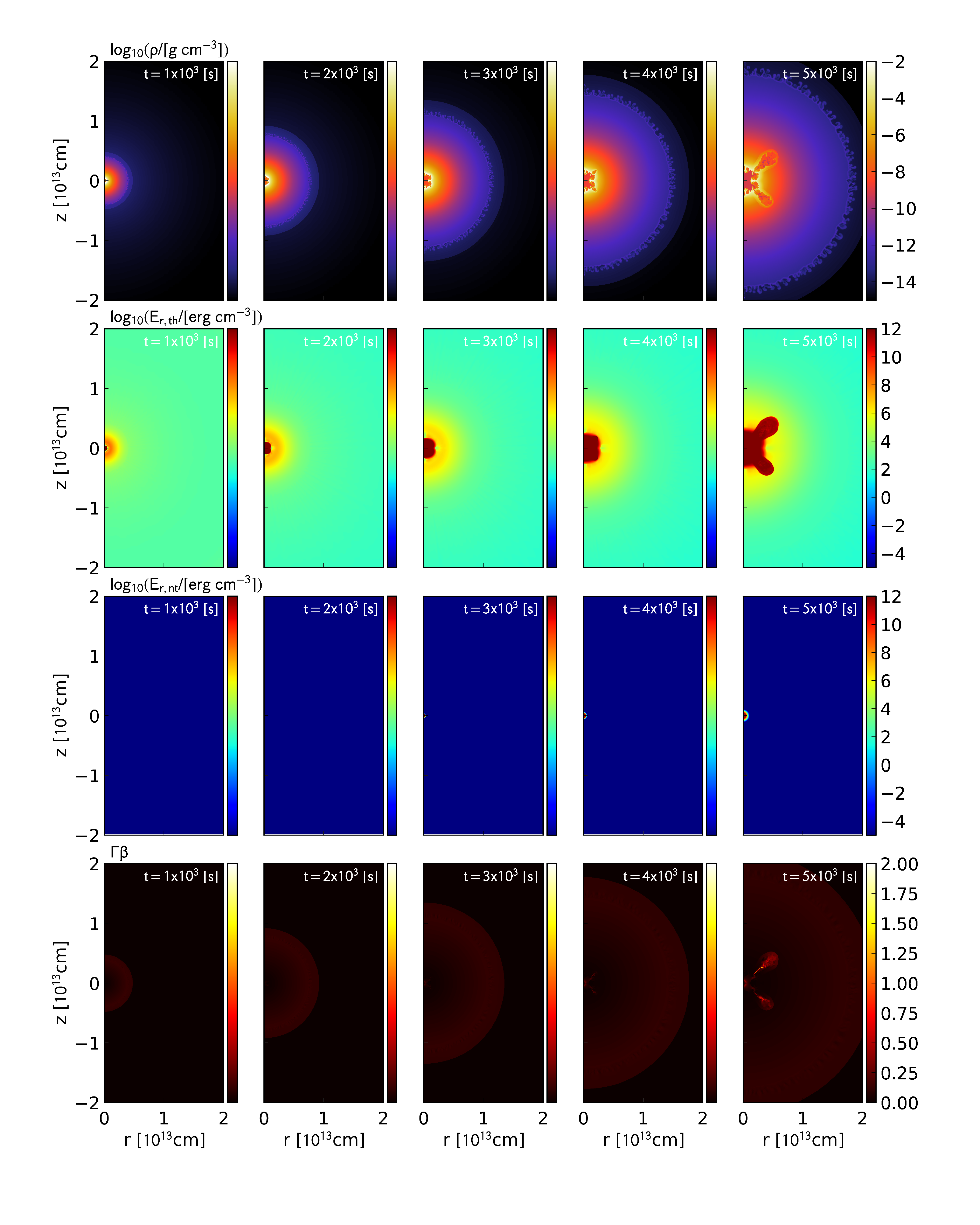}
\cprotect\caption{Results of model \verb|L48large|. Spatial distributions of density, gas energy density, radiation energy density, and 4-velocity are plotted from top to bottom. 
The columns present distributions at $t=10^3$, $2\times10^3$, $3\times10^3$, $4\times10^3$, and $5\times10^3$ s from left to right. }
\label{fig:evol1_L48large}
\end{center}
\end{figure*}
\begin{figure*}
\begin{center}
\includegraphics[scale=0.4]{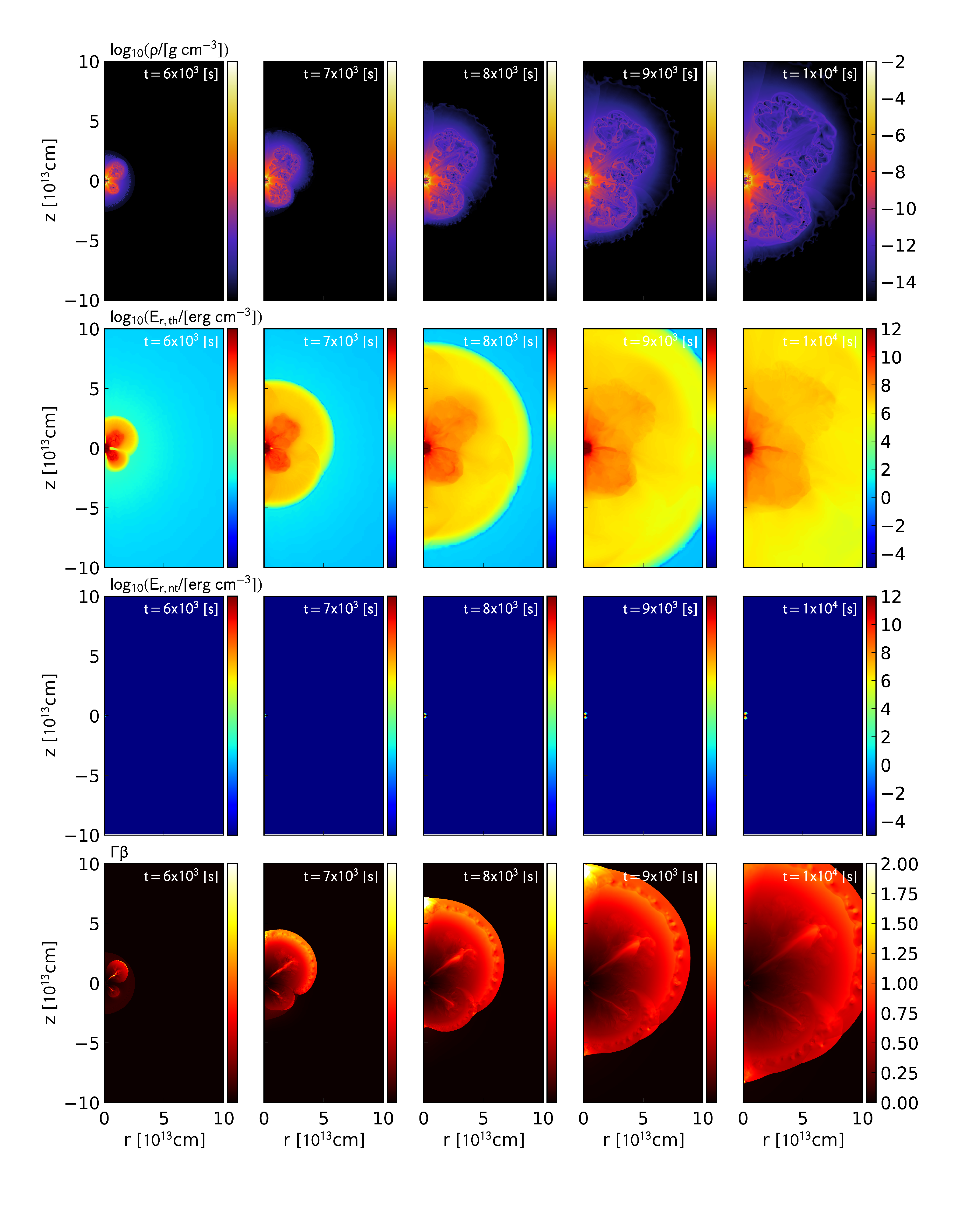}
\cprotect\caption{Same as Figure \ref{fig:evol1_L48large}, but at $t=6\times10^3$, $7\times10^3$, $8\times10^3$, $9\times10^3$, and $10^4$ s from left to right. }
\label{fig:evol2_L48large}
\end{center}
\end{figure*}

\subsubsection{Temporal evolution}
The SN ejecta with the central energy source with $L_\mathrm{in,0}=10^{48}$ erg s$^{-1}$ evolves in a shorter timescale than those with the lower energy injection rate, because of the short characteristic timescale of the energy injection, $E_\mathrm{sn}/L_\mathrm{in,0}=10^3$ s. 
Even for the most energetic model \verb|L48large|, the energy injection rate starts declining rapidly at $t_\mathrm{in}=10^4$ s, which is much shorter than the typical photon diffusion timescale in the ejecta, Equation \ref{eq:t_diff}. 
In other words, photons are tightly coupled with gas in most parts of the ejecta, which distinguishes this model series from that with $L_\mathrm{in,0}=10^{46}$ erg s$^{-1}$ shown above. 
Therefore, it is naturally expected that the resultant dynamical evolution of the ejecta is more similar to the previous hydrodynamic simulations without radiative transfer effects. 

The dynamical evolution of model \verb|L48large| shown in Figures \ref{fig:evol1_L48large} and \ref{fig:evol2_L48large} is indeed similar to purely hydrodynamic cases. 
The expansion of the quasi-spherical hot bubble, the development of the Rayleigh-Taylor instability, and the shock breakout driven by the hot bubble pressure are seen. 
The epoch of the runaway acceleration of the forward shock is in agreement with the theoretical estimate of $t\simeq 5t_\mathrm{c}(=5\times 10^3\ \mathrm{s})$ \citep{2017MNRAS.466.2633S}, which is based on a self-similar solution and confirmed by hydrodynamic simulations. 

One important difference from the models with the lower energy injection rate is that the non-thermal radiation is well confined in the central region even at $t>t_\mathrm{in}$. 
This is simply because the surrounding ejecta is still dense and opaque to non-thermal photons. 
The additional energy injected in the form of non-thermal radiation is almost immediately thermalized. 
The hot bubble breakout surely produces filamentary density structure in this case. 
However, even at the cavities created by the hot bubble breakout, the density is high enough to trap non-thermal photons. 
The non-thermal photon leakage occurs at significantly later epochs in this model, when the energy injection rate drops to values much smaller than the initial rate of $L_\mathrm{in,0}=10^{48}$ erg s$^{-1}$. 
As a result, the fraction of the escaped non-thermal radiation energy out of the total injected energy is significantly smaller as indicated in Figures \ref{fig:energy_L46} and \ref{fig:energy_L48} below. 

Another remarkable feature is the efficient acceleration of the outermost layer. 
When the forward shock driven by the hot bubble emerges from the surface of the ejecta, the outer layers of the ejecta are accelerated to relativistic velocities, $\Gamma\beta>1$ as seen in Figure \ref{fig:evol2_L48large} (bottom panels). 
This is in contrast to the models with the lower energy injection rate, in which the outer layers are expanding only at sub-relativistic velocities, $\Gamma\beta\sim 0.1$. 
This difference emphasizes the importance of the injection timescale in producing the relativistic ejecta component. 
As we have seen in the previous subsection, the injection timescale closer to the photon diffusion timescale enhances the conversion efficiency of the injected energy into thermal radiation. 
Instead, such long injection timescales make it inefficient to accelerate the outer envelope due to the weak coupling between gas and radiation in the outer layer. 
On the other hand, for the models with the high energy injection rate, the central energy injection happens while even the outer envelope of the ejecta is opaque to thermal and non-thermal photons. 
Therefore, the forward shock powered by the hot bubble does not suffer from a significant radiative loss and successfully deposits the shock energy into a smaller and smaller amount of gas as it propagates in a more and more dilute part of the ejecta. 

\begin{figure*}
\begin{center}
\includegraphics[scale=0.5]{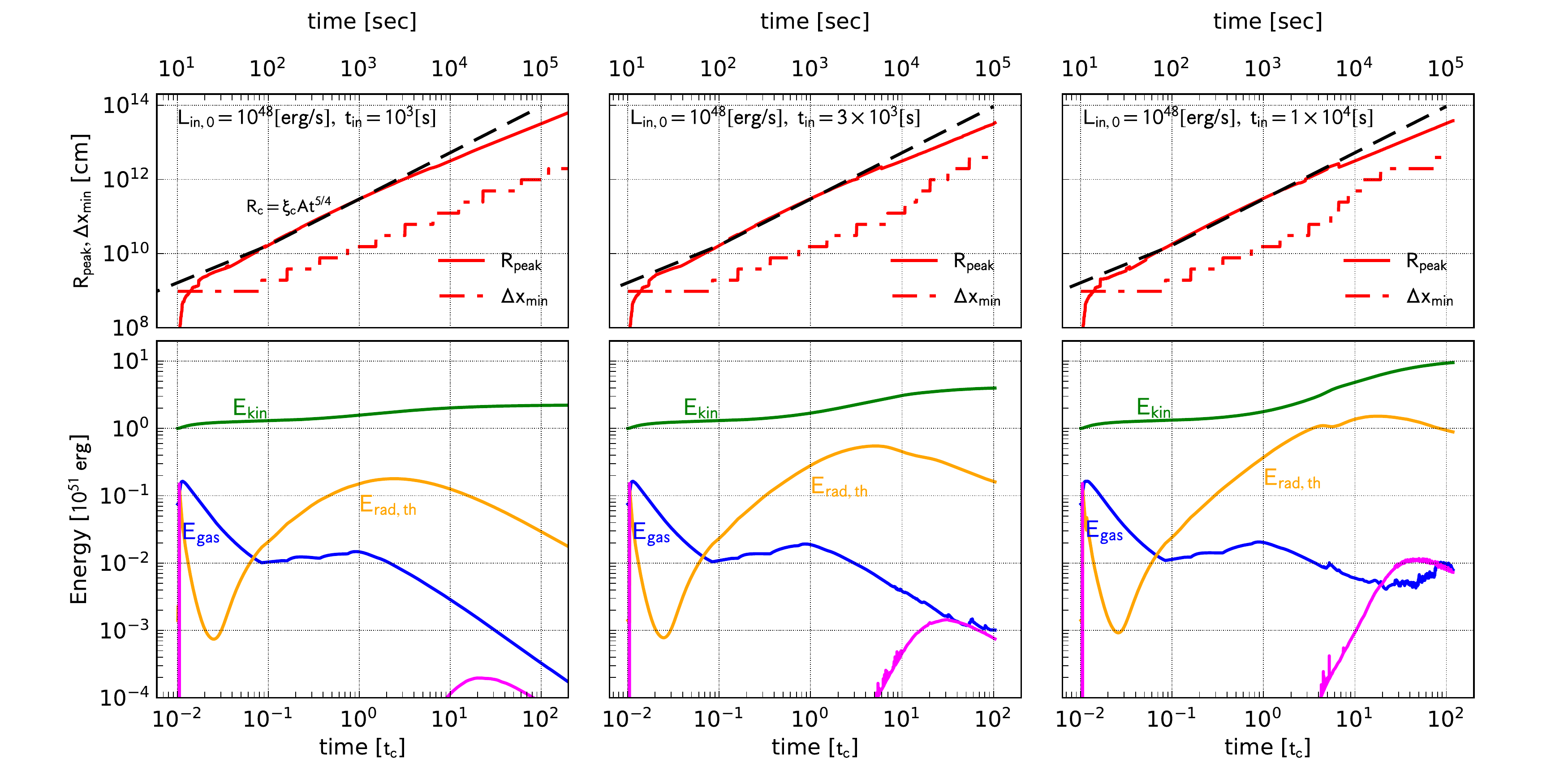}
\caption{Same as Figure \ref{fig:energy_L46}, but for the models with $L_\mathrm{in,0}=10^{48}$ erg s$^{-1}$. }
\label{fig:energy_L48}
\end{center}
\end{figure*}

Figure \ref{fig:energy_L48} shows the temporal evolutions of the radius of the density peak and the kinetic, gas internal, thermal radiation, and non-thermal radiation energies. 
The overall trends of these quantities are similar to the \verb|L46| model series. 
The self-similar evolution of the density peak, $R_\mathrm{max}$, and its deviation from the self-similar law at later epochs are reproduced. 
The gas internal energy again plays a minor role in the dynamical evolution. 
Although the internal energy is dominated by the thermal radiation energy, which becomes as large as $10\%$ of the total energy, thermal radiation is mostly trapped in the expanding ejecta and would decrease according to adiabatic cooling. 
The non-thermal radiation energy is much smaller compared with the \verb|L46| model series, because of the efficient absorption of non-thermal photons by the high-density ejecta.

\subsubsection{Kinetic energy distribution}
The kinetic energy distribution of the ejecta gives a more quantitative look at this efficient acceleration of the outer layers.  
As is usually done to characterize the outer envelope of SN ejecta \citep[e.g.][]{1999ApJ...510..379M,2001ApJ...551..946T}, we calculate the mass and the kinetic energy of the ejecta component traveling at 4-velocities higher than a given value $\Gamma\beta$,
\begin{equation}
    M(\Gamma\beta)=2\pi \int_{>\Gamma\beta}\rho\Gamma rdrdz,
\end{equation}
and
\begin{equation}
    E_\mathrm{k}(\Gamma\beta)=2\pi \int_{>\Gamma\beta}\rho\Gamma(\Gamma-1) rdrdz,
\end{equation}
where the integrations are carried out only for numerical cells with the local 4-velocity greater than $\Gamma\beta$. 

\begin{figure*}
\begin{center}
\includegraphics[scale=0.6]{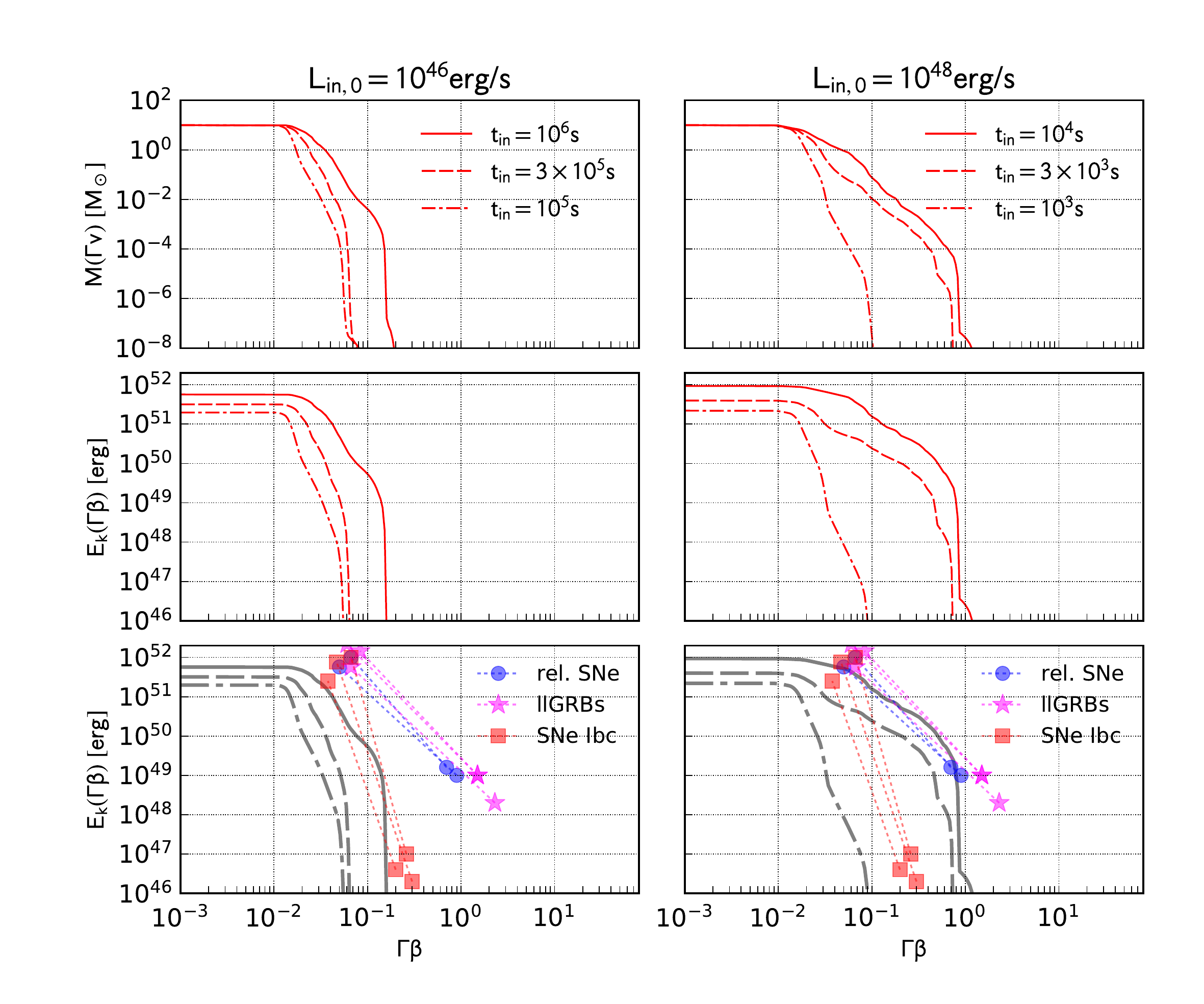}
\caption{Mass and kinetic energy distributions of the ejecta having experienced the central energy injection. 
The mass (top) and the kinetic energy (middle) of the ejecta component traveling faster than the 4-velocity $\Gamma\beta$ are plotted. 
The left and right columns correspond to the models with $L_\mathrm{in,0}=10^{46}$ and $10^{48}$ erg s$^{-1}$, respectively. 
In the bottom panels, the numerical kinetic energy distributions are compared with observations of several SNe (blue circle: relativistic SNe, magenta star: SNe with low-luminosity GRBs, and red square: Type-Ibc SNe). 
The plotted objects and their references are as follows; 2009bb \citep{2013MNRAS.434.1098C}, 
2012ap \citep{2014ApJ...797..107M,2015ApJ...799...51M}, 
1998bw/GRB980425 \citep{1998Natur.395..663K,2013MNRAS.434.1098C},
2006aj/GRB060218 \citep{2013ApJ...778...18M,2013MNRAS.434.1098C},
2010bh/GRB100316D \citep{2013ApJ...778...18M,2013MNRAS.434.1098C},
2002ap \citep{2003ApJ...599..408B,2013MNRAS.434.1098C},
2007gr \citep{2010ApJ...725..922S},
2012au \citep{2014ApJ...797....2K,2013ApJ...770L..38M}}
\label{fig:spec}
\end{center}
\end{figure*}

Figure \ref{fig:spec} shows the mass and the kinetic energy distributions for all the models. 
The distributions for the \verb|L46| model series do not extend up to $\Gamma\beta\simeq 1$. 
As we have discussed in Section \ref{sec:dynamical_evolution}, this again highlights the difference from the previous hydrodynamic simulations in 2D and 3D \citep{2017MNRAS.466.2633S,2019ApJ...880..150S}. 
On the other hand, the models with the high energy injection rate and large injected energies, namely models \verb|L48large| and \verb|L48mid|, reproduce trans-relativistic ejecta components traveling at velocities as fast as $\Gamma\beta\simeq 1$, which is in agreement with the previous hydrodynamic simulations. 
This agreement reassures the idea that the purely hydrodynamic simulations represent the optically thick limit of radiation-hydrodynamic simulations. 

The outer envelope of SN ejecta drives a blast wave propagating in the circumstellar medium. 
The blast wave sweeps the ambient medium and produces non-thermal electrons via diffusive shock acceleration, which outshine via synchrotron emission. 
The more energetic the expanding outer envelope is, the brighter the associated synchrotron emission becomes. 
This is how the radio synchrotron emission observed from some SNe is used to estimate how much energy is distributed in the outermost layer of the SN ejecta. 
Combined with optical observations, which probe the dynamics of the optically thick part of the ejecta, radio synchrotron luminosity therefore constrains the kinetic energy distribution. 
In the bottom panels of Figure \ref{fig:spec}, we plot kinetic energies of the outer envelope and the optically thick ejecta of some SNe in the literature (see the figure caption for the plotted objects and their references).
Remarkably, the model \verb|L48large| can explain the properties of the photospheric and radio synchrotron emission from radio-bright SNe Ic-BL, such as SN 2009bb, thanks to the efficient acceleration of the outer layer, while the models with the lower energy injection rate cannot. 
The implications of this finding on the diversity of energetic SNe will be discussed in further detail in Section \ref{sec:discussion}. 

\subsubsection{Radial distributions}
\begin{figure*}
\begin{center}
\includegraphics[scale=0.5]{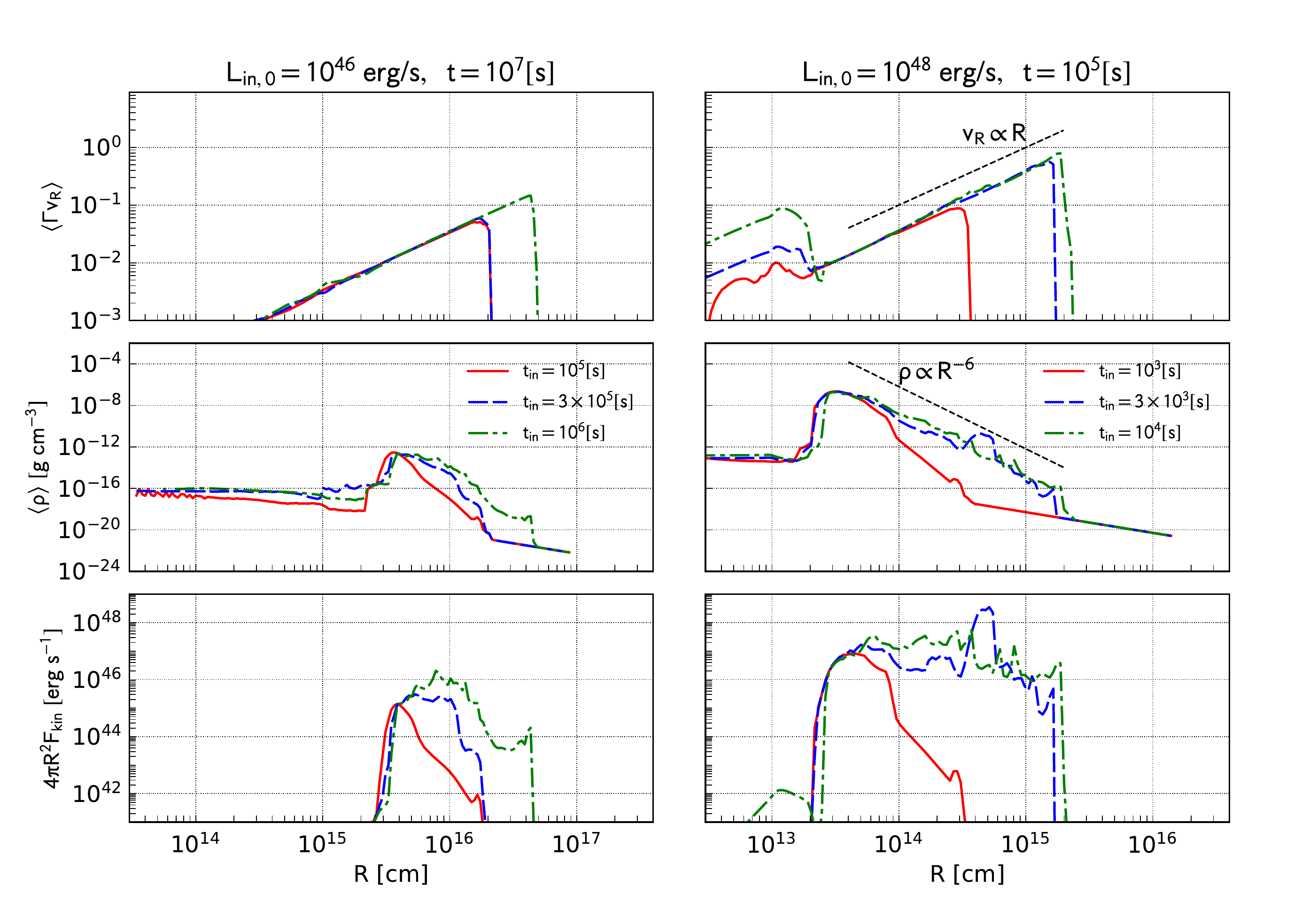}
\cprotect\caption{Angular-averaged radial distributions of 4-velocity (top), density (middle), and outgoing kinetic luminosity (bottom) at $t=10^2t_\mathrm{c}$. 
}
\label{fig:radial}
\end{center}
\end{figure*}
As in \cite{2017MNRAS.466.2633S}, we calculate the radial distribution of a hydrodynamic variable $Q$ by the following mass-weighted angular average,
\begin{equation}
    \langle Q(R)\rangle=\frac{2\pi}{\Delta M}\int^{R+\Delta R/2}_{R-\Delta R/2}
    \int^{\pi}_{0}Q\rho \Gamma \sin\theta rdrd\theta
\end{equation}
with 
\begin{equation}
    \Delta M=2\pi\int^{R+\Delta R/2}_{R-\Delta R/2}
    \int^{\pi}_{0}\rho \Gamma \sin\theta rdrd\theta
\end{equation}
where the integrations are carried out for numerical cells in the annulus, $R-\Delta R/2\leq R<R+\Delta R/2$. 

In Figure \ref{fig:radial}, we compare the radial distributions of the 4-velocity, the density, and the outgoing kinetic energy for the models considered in this work. 
The radial distributions are obtained at $t=10^2 t_\mathrm{c}$ for both the \verb|L46| and \verb|L48| model series. 
For all the models, the radial velocity is well proportional to the radius, $v_\mathrm{R}\propto r$, indicating that the ejecta is in the free expansion phase. 
The common feature in the radial density distribution is the presence of the central cavity, which is created by the central energy injection. 
On the other hand, the outer density structure is different among the models.   
The radial density distributions of the models \verb|L48mid| and \verb|L48large| (the right panel) are well described by a power-law function with an index $-6$, $\rho\propto r^{-6}\propto v_\mathrm{R}^{-6}$. 
This is in agreement with the previous 2D hydrodynamic simulation \citep{2017MNRAS.466.2633S}. 
The efficient energy transport from the central energy source to outer layers realizes relatively shallow density and almost flat kinetic energy profiles. 
In \cite{2017MNRAS.466.2633S}, we analytically derived the power-law density profile with the index between $-5$ and $-6$ for a flat energy spectrum. 
Therefore, these radial distributions of the density and the outgoing kinetic energy are also in good agreement with the analytical consideration. 
For the model with the smallest injected energy, \verb|L48small|, only the inner part of the ejecta is affected by the energy injection. 
Therefore, the outer density profile remains steep as initially assumed. 

For the lower energy injection rate $L_\mathrm{in,0}=10^{46}$ erg s$^{-1}$, the density profiles are not well described by a power-law function even in the free expansion phase. 
As we have seen in previous sections, the radiative loss prevents efficient energy transport from the center to the outer layers. 
Therefore, only a smaller amount of energy is distributed to the outer layers of the ejecta.

\section{Discussion\label{sec:discussion}}
\subsection{Effects of radiative transfer}
As we have seen in the previous section, the effects of radiative transport are of critical importance to determine the dynamical evolution of SN ejecta with a powerful central energy source and the resultant emission properties. 
In the following, we summarise the important effects of radiative diffusion.

\subsubsection{Radiative efficiency}
First of all, an energy injection timescale a bit shorter ($t_\mathrm{in}=0.01$--$0.1t_\mathrm{diff}$) than the photon diffusion timescale makes the thermal emission highly luminous. 
As demonstrated by the models with $L_\mathrm{in,0}=10^{46}$ erg s$^{-1}$, the injection timescales of a few days to 10 days realize a peak bolometric luminosity brighter than $10^{45}$ erg s$^{-1}$. 
Even the model with the total injected energy of $10^{51}$ erg produces thermal emission as bright as $5\times 10^{43}$ erg s$^{-1}$. 
The total radiated energy for each model is a considerable fraction of the injected energy. 

On the other hand, the models with $L_\mathrm{in,0}=10^{48}$ erg s$^{-1}$ assume the energy injection timescale much shorter than the photon diffusion timescale. 
For these models, we expect no bright thermal emission lasting for $>$10 days (see, Figure \ref{fig:energy_L48}). 
We stopped simulations at a few $10^{5}$ s (a few days), at which the thermal radiation energy less than $\sim 10^{51}$ erg is still trapped in the ejecta since the most part of the ejecta is optically thick. 
This thermal energy suffers from adiabatic cooling as it diffuses through the ejecta in $\sim 10$ days, significantly reducing the radiative efficiency. 
Therefore, these models with $L_\mathrm{in,0}=10^{48}$ erg s$^{-1}$ are unlikely to explain SLSNe-I with their total radiated energies as much as $10^{51}$ erg. 
Their thermal emission at several 10 days is powered by the radioactive decay of freshly synthesized $^{56}$Ni and they are likely observed as SNe with high expansion velocities, i.e., SNe Ic-BL. 

\subsubsection{Bright thermal emission}
Bright thermal emission is observed for the models with $L_\mathrm{in,0}=10^{46}$ erg s$^{-1}$. 
We emphasize that the light curves obtained in these models behave differently from the 1D counterparts as shown in Figure \ref{fig:lc_L46}. 
It is widely known that SN light curves are dependent on the density structure and the distribution of the energy source. 
In the case of SNe with a powerful central energy source, the central energy injection itself can modify the ejecta density and velocity structure. 
In addition, the modified density structure also modifies the mean free path of non-thermal radiation and thus the way of the energy deposition into the SN ejecta. 
As a result, the light curves of both thermal and non-thermal radiation evolve more rapidly in 2D models than those in the 1D counterparts. 
We note that the still uncertain non-thermal opacity should also influence the radiative efficiency and the light curve. 

This difference suggests that the interpretation of the observed light curves of SLSNe in terms of a one-zone or 1D radiation hydrodynamic model should be done carefully. 
As seen in Figure \ref{fig:lc_L46}, the light curves basically follow the energy injection rate at the late epochs. 
Therefore, the rate and the timescale of the energy injection can be well constrained from observed light curves. 
On the other hand, the peak luminosity and the with of the main bump strongly depend on the ejecta structure. 
In particular, it is expected that one-zone and 1D light curve models predict a longer evolutionary timescale, i.e., a larger ejecta mass and/or lower velocity for the same parameter set. 
Such variables estimated by simple light curve models may involve systematic offsets from the true values.

\subsubsection{Sub-relativistic outermost layer}
The injected energy is also used for accelerating outer layers of the ejecta in the models with the higher injection rate. 
As demonstrated in Figure \ref{fig:spec}, the kinetic energy spectra of these models show the presence of relativistic ejecta traveling at velocities close to the speed of light. 
When the forward shock driven by the hot bubble propagates in the outer ejecta, it is the radiation pressure that pushes the shock front. 
In the models with the higher energy injection rate, the tight coupling between gas and radiation in the outer ejecta realizes the energy deposition and acceleration by the strong shock. 
On the other hand, in the models with the lower energy injection rate, the outer ejecta is only loosely coupled with radiation when they are hit by the forward shock. 
Then, the radiation in the shocked region diffuses away rather than efficiently energizing the gas in the outer layers.

\subsection{Impact of the hot bubble breakout}
The total amount of the injected energy is another important parameter. 
When the injected energy is a few times more than the initial kinetic energy of the ejecta, the inflating hot bubble blows away the ejecta, resulting in a number of characteristic features in the ejecta structure and the expected emissions. 
Even when the injected energy is comparable to the initial kinetic energy, the inner ejecta is inevitably affected by the material mixing.

\subsubsection{Breakout emission}
When the forward shock driven by the hot bubble successfully emerges from the photosphere in the outer ejecta, the radiation energy having been trapped in the post-shock region starts escaping into the surrounding space. 
This is in analogy to the shock breakout emission from the stellar surface in normal SN explosions. 
The shock breakout emission from the hot bubble is seen as an initial spike in the bolometric light curve of the  model \verb| L46large| (Figure \ref{fig:lc_L46}). 
This is in contrast to the 1D spherical simulations with smoothly evolving light curves. 
Assuming that the photospheric radius and the bolometric luminosity at the breakout is of the order of $10^{15}$ cm and $10^{45}$ erg s$^{-1}$, the effective temperature is estimated to be $T_\mathrm{eff}\simeq 3$--$4\times 10^4$ K. 
Therefore, we expect UV-dominated bright emission at the beginning. 
This initial bright emission differentiates itself from the subsequent emission powered by the continued energy supply and may correspond to the early bumps in some SLSN light curves \citep{2012A&A...541A.129L,2015ApJ...807L..18N,2017ApJ...835...58V}. 

We note that the properties of the breakout emission would be highly dependent on how the Rayleigh-Taylor fingers and the associated fast outflows from the hot bubble develop.  
One caveat on our 2D simulations is the artificial bipolar structure produced due to the assumed axi-symmetry. 
As a result, the large-scale shock front penetrates the outer ejecta almost simultaneously, making the shock breakout emission prominent. 
In our previous 3D hydrodynamic simulation \citep{2019ApJ...870...38S}, however, the shocked region consists of a number of Rayleigh-Taylor fingers rather than showing a global bipolar structure. 
Therefore, in a more realistic 3D setting, the hot bubble breakout may be moderate, thereby making the associated breakout emission less bright.

\subsubsection{Light curve undulation}
The hot bubble breakout also has an impact on the light curve in the late epochs. 
As we have seen in Figure \ref{fig:lc_L46}, the light curve of the model \verb|L46large| exhibits slightly more complicated time variability in the decaying phase than the model \verb|L46small|, which shows a rather smooth light curve. 
This is likely due to the different impact of the hot bubble breakout. 
In the model \verb|L46small|, the hot bubble is still contained by the SN ejecta at several 10 days (Figure \ref{fig:comparison_L46}). 
This results in the smooth light curve with less significant viewing angle dependence, which is also expected from the smooth spatial distribution of the thermal radiation energy density in Figure \ref{fig:comparison_L46} (the bottom panel). 
In the model \verb|L46large|, on the other hand, the photosphere recedes into the layers having been affected by the hot bubble while outshining in bright thermal emission. 
This makes the thermal radiation field highly anisotropic as seen in the middle and bottom panels of Figure \ref{fig:comparison_L46}. 
The resultant light curve is highly dependent on the line of sight. 

Interestingly, one of the intriguing features in the light curves of SLSNe-I is the so-called undulation, fluctuations in the measured magnitudes \citep{2016ApJ...826...39N,2017MNRAS.468.4642I,2018ApJ...865....9B}. 
The photosphere disturbed by the hot bubble breakout is a promising way to explain this peculiar light curve behavior in SLSNe-I, although other explanations, such as CSM interaction, cannot be excluded. 
If the light curve undulation is actually an imprint of the ejecta destruction by the hot bubble, associated spectral changes would be expected. 
The inner layers of the ejecta have been swept by the hot bubble and form a shell with a number of low-density channels penetrated by the fast outflows from the hot bubble. 
When the receding photosphere reaches the shell, the ejecta becomes transparent to photons for lines of sight penetrating the low-density channels. 
Therefore, the presence of the undulation would at least partially coincide with the spectral transition from the photospheric to the nebular phase.

\subsection{Non-thermal emission}
Highly energetic SNe, which may harbor a central engine, are often associated with bright synchrotron and inverse Compton emission in radio and X-ray, which serves as an important probe for the physical mechanism making them energetic. 

\subsubsection{Non-thermal radiation leakage}
In the central engine scenarios for SLSNe-I, the non-thermal emission from the central compact object or the associated wind nebula is the power source for the bright emission. 
Therefore, direct detection of the long-lasting non-thermal emission from the central engine can be a smoking gun to reveal the physical association of an SLSN with such a power source \citep[e.g.,][]{2014MNRAS.437..703M,2015ApJ...805...82M,2016MNRAS.461.1498M,2016ApJ...818...94K}. 
Follow-up radio and X-ray observations of SLSNe-I from several 10 days to a few years after the discovery have been attempted, however, resulting in no undoubtful detection so far \citep{2013ApJ...771..136L,2018ApJ...856...56C,2018ApJ...864...45M,2018A&A...611A..45R,2018ApJ...868L..32B}. 
Very recently, a bright radio source was identified in association with an almost 10-years old SLSN, PTF10hgi \citep{2019ApJ...886...24L}, which may correspond to the theoretically predicted non-thermal radio emission from the embedded magnetized neutron star \citep[e.g.,][]{2018MNRAS.474..573O}. 

The SN ejecta surrounding the putative central engine prevents both high-energy photons and radio waves from escaping into outer space and thus the ejecta density structure and the expansion velocity play a decisive role in determining the detection feasibility. 
The hot bubble breakout is again important for the escape probability of non-thermal photons. 
As shown in Figure \ref{fig:lc_L46}, our simulations demonstrate that the non-thermal radiation starts leaking from the ejecta much earlier in multi-dimensional setting than in 1D spherical cases. 
The non-thermal radiation leakage would also be closely related to the missing energy problem \citep{2018ApJ...868L..32B}. 
The leakage is more effective with a larger injected energy because a more powerful engine blows away the surrounding ejecta more effectively. 
The shredded SN ejecta has low-density channels with low plasma frequency, which allow radio waves easily transmit in the late epochs.

\subsubsection{Blast wave emission}
For models with the higher energy injection rate, a sufficiently large injection energy produces outer layers accelerated to relativistic velocities. 
The blast wave driven by the relativistic ejecta accelerates electrons as it sweeps the surrounding gas. 
The synchrotron and the inverse Compton emission from non-thermal electrons are also probed by radio and X-ray observations and serve as an important clue to how much energy is deposited in the outermost layer of the SN ejecta \citep[e.g.,][]{2006ApJ...651..381C}. 

In our previous work \citep{2018MNRAS.478..110S}, we studied broad-band emission properties of SN ejecta with the density and velocity structure implied by our previous 2D hydrodynamic simulation \citep{2017MNRAS.466.2633S} and compared the resultant radio and X-ray light curves with observations of energetic and/or luminous SNe. 
We found that the relativistic outer layers indeed produce bright radio and X-ray emission, which potentially explains radio-bright SNe, such as SN 1998bw and 2009bb. 
On the other hand, however, the radio non-detection of the SLSN-I 2015bn \citep{2016ApJ...826...39N}, 2017egm \citep{2018ApJ...853...57B}, and Gaia16apd \citep{2018ApJ...856...56C} indicates that the efficient acceleration of the outer layers may not have happened in these well-studied SLSNe-I. 

Our radiation-hydrodynamic simulations explain the difference. 
The models with the lower energy injection rate produce thermal emission as luminous as SLSNe-I, while they fail to accelerate outer layers up to relativistic speeds due to the leakage of thermal radiation. 
On the other hand, the models with the higher energy injection rate are less likely to explain the extreme brightness of SLSNe-I, while the energy injection instead creates energetic ejecta with outer layers traveling at relativistic velocities. 
The former case is in agreement with SLSNe-I without bright radio emission and the latter case explains energetic SNe with bright radio counterparts, such as SN 1998bw and 2009bb. 
As pointed out by \cite{2018MNRAS.478..110S}, the power-law density profile with the slope of $-6$ is in good agreement with the analysis of the radio-bright SN 2012ap \citep{2015ApJ...805..187C}, suggesting that the hot bubble breakout and the subsequent energy transport from the central engine to the outer layers is a promising way to explain the origin of radio-bright SNe Ic-BL.

\subsection{Ejecta structure probed by spectroscopic and polarimetric observations}

\subsubsection{Spectra in the photospheric phase}

The spectra of SLSNe-I in the photospheric phase probe the density and chemical stratification. 
Our simulation results imply that the outer ejecta evolves differently depending on whether it is blown away by the central hot bubble or not. 
In the case of the violent hot bubble breakout, the outer ejecta is shredded and the density distribution is flattened. 
The spectra in the photospheric phase would be the superposition of the emission/absorption lines from a wide variety of velocity components. 
As a result, the spectra would look broad-lined. 
On the other hand, a smaller injected energy leaves the outer layers well-stratified. 
Although the ionization structure of the outer layer can be different from those of normal SNe due to the continuous heating by the diffusing photons from the center, the outer density and velocity structure would be similar to normal SNe. 
Therefore, the spectra of SLSNe-I with a moderately powerful central engine may look similar to normal SNe in terms of absorption/emission line width. 

\cite{2018ApJ...855....2Q} recently analyzed the spectra of 19 SLSNe-I from the Palomar Transient Factory (PTF). 
They classified their sample into two groups, SN 2011kl-like or PTF12dam-like objects, in terms of their spectral similarity. 
The former objects exhibit  smoother spectra and shorter declining timescales than the latter objects. 
This spectral diversity among observed SLSNe-I may be partly due to the different impact of the embedded power source on the SN ejecta. 

The material mixing by the hot bubble breakout is again important for considering how and when spectral features of different elements emerge. 
The gas flows penetrating the SN ejecta transport elements synthesized in the deep interior of the ejecta toward outer layers (Figure \ref{fig:mixing}). 
This results in absorption lines of such heavy elements appearing in the early stages of the spectral evolution. 
Recently, observations of SLSN-I 2017dwh around the optical maximum \citep{2019ApJ...872...90B} have identified a strong absorption line likely caused by \ion{Co}{2}. 
This ``chemical inversion'' could be the evidence of the inner engine activity and in agreement with our results. 
We also note that the ``chemical inversion'' has been found in the (low-luminosity) gamma-ray bursts GRB 171205A/SN 2017iuk and 161219B/SN 2016jca, which are likely powered by a jet activity \citep{2019Natur.565..324I,2019MNRAS.487.5824A}. 

We note that there are several other ways to make the stratified layers mixed both in the pre-supernova stage and the first explosion, although the hot bubble breakout is an efficient way to achieve inverted chemical structures. 
The turbulent convection in massive stars at the pre-supernova stage has been investigated by multi-dimensional hydrodynamic simulations \citep[e.g.][]{2007ApJ...667..448M,2011ApJ...733...78A,2016RPPh...79j2901A,2019ApJ...881...16Y}, which revealed that the convective motion plays important roles in making the inner stellar structure inhomogeneous. 
The following neutrino-driven convection in the explosion phase can make the ejecta inhomogeneous \citep[e.g.,][]{2000ApJ...531L.123K,2003A&A...408..621K,2006A&A...453..661K}. 
The shock passage through the stratified layers with different chemical compositions is also known to cause the Rayleigh-Taylor mixing at the interface of the layers \citep{1989ApJ...341L..63A,1990ApJ...358L..57H}. 
These processes would also contribute to the chemical structure of SN ejecta even without the hot bubble breakout. 

\subsubsection{Nebular spectra}

The inner density structure of the ejecta can be probed by optical spectroscopic observations when the entire ejecta becomes transparent to optical photons, i.e., in the nebular phase. 
Because of the modification of the density structure and the inner mixing, the nebular spectra of SN ejecta with a central power source would show some remarkable features. 
There have been a number of observational studies on the nebular emission from SLSNe-I \citep{2009Natur.462..624G,2015MNRAS.452.1567C,2015ApJ...814..108Y,2016ApJ...831..144L,2016ApJ...828L..18N,2019ApJ...871..102N}. 
SLSNe-I show nebular phase spectra similar to those of SNe Ic-BL, implying a physical link between them. 

\cite{2017ApJ...835...13J} applied their theoretical nebular emission model for some well-observed SLSNe-I. 
Their analysis suggests that a massive oxygen-rich layer ($>10M_\odot$) and clumpy emitting regions with a small filling factor of $<0.01$ are required to explain the bright oxygen and magnesium nebular lines. 
Moreover, the high calcium line ratio implies a rather high electron density, $\gtrsim 10^8$ cm$^{-3}$. 
Interestingly, clumpy inner ejecta is naturally produced by the hot bubble expansion. 
As we have demonstrated in Figure \ref{fig:dVdlog_n}, the highest ion number density is $n_\mathrm{i}=10^8$--$10^9$ cm$^{-3}$. 
Thus, the electron number density as high as $10^{8}$ cm$^{-3}$ can easily be achieved by considering singly or doubly ionized oxygen-rich ejecta. 
The filling factor of the region with the highest ion number density is as small as $0.01$, which also agrees with the analysis of \cite{2017ApJ...835...13J}. 
This finding suggests that the clumpy inner layer produced by our simulations is in a remarkable agreement with the spectral features of SLSNe-I in their nebular phase, although detailed line transfer calculations based on the simulation results are required for fair comparisons of the models with observations.

\subsubsection{Inner ejecta structure probed by polarimetric observations}

Polarization signals from SLSNe-I are also an important probe for the multi-dimensional ejecta structure \citep[e.g.,][]{2015ApJ...815L..10L,2017ApJ...837L..14L,2016ApJ...831...79I,2018ApJ...853...57B,2018MNRAS.479.4984C,2019ApJ...875..121L,2019MNRAS.482.4057M,2020ApJ...894..154S}. 
Polarimetric observations of some SLSNe-I reported no significant linear polarization around the optical maximum (e.g, LSQ14mo; \citealt{2015ApJ...815L..10L}, 2018hti; \citealt{2019ApJ...875..121L}), suggesting a nearly spherical photosphere. 
On the other hand, SN 2015bn \citep{2016ApJ...831...79I,2017ApJ...837L..14L} showed detectable linear polarization in multiple epochs. 
The increased polarization degree from the first to the second observing epochs indicates the presence of the aspherical inner core, which is revealed as the photosphere recedes in the ejecta.  
The nearby SLSN-I 2017egm showed a similar increase in the polarization degree \citep{2018ApJ...853...57B,2019MNRAS.482.4057M,2020ApJ...894..154S}. 
\cite{2017ApJ...837L..14L} point out that the increase in the polarization degree in SN 2015bn coincides with the spectral transition. 
If the central engine model for SLSNe-I is the case, the radial structure of the ejecta is divided into the inner part affected by the violent mixing and the well-stratified outer part as demonstrated by our simulations. 
The inner part is more sensitive to the geometry of the hot bubble, which could be aspherical. 
This picture may be consistent with the evolution of SLSNe-I revealed by spectropolarimetric observations, although more objects with spectropolarimetirc data should be accumulated in future studies.

\subsection{A unified scenario for energetic supernovae} 
\begin{figure*}
\begin{center}
\includegraphics[scale=0.6]{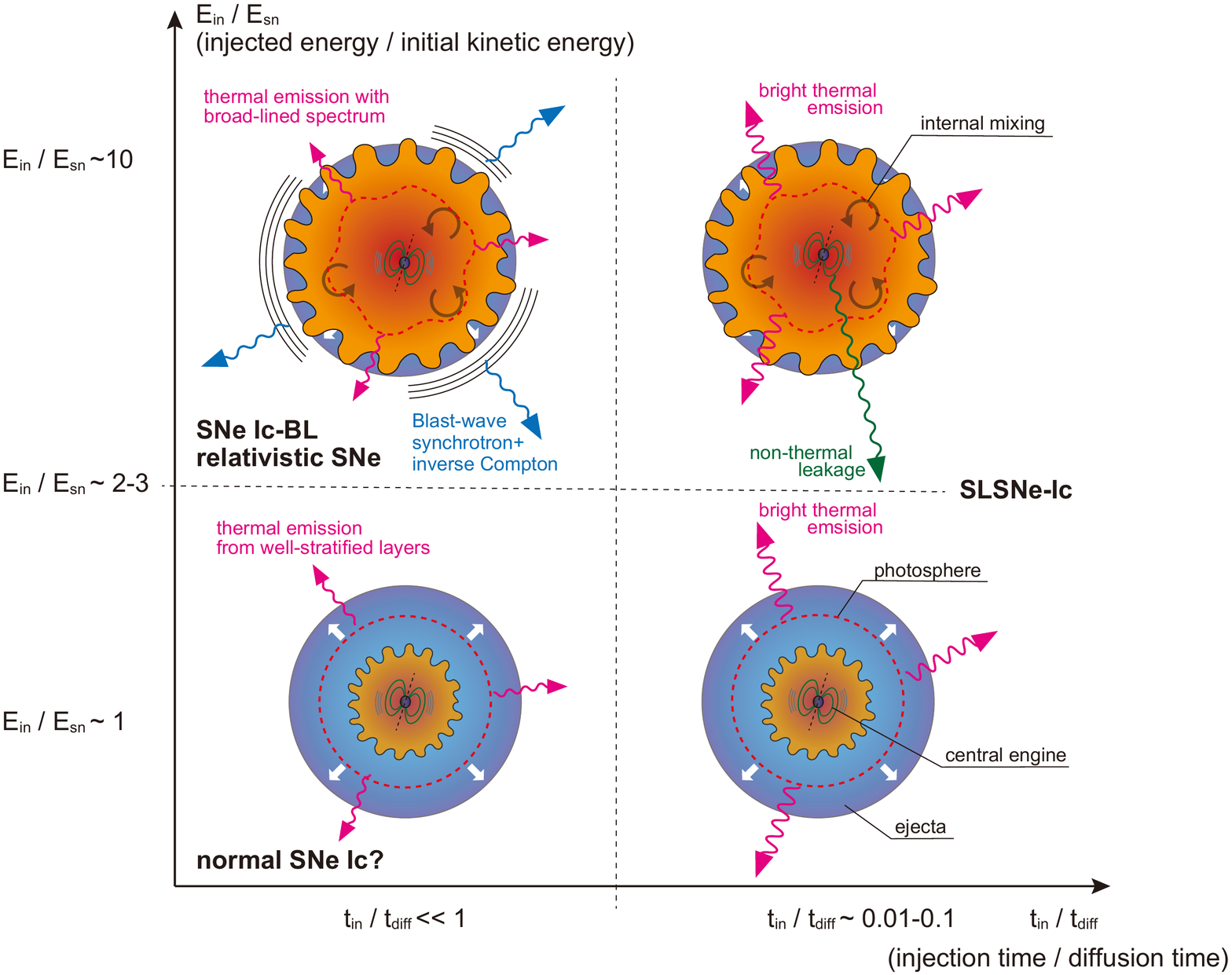}
\caption{Schematic representation of the parameter space of SNe with a central power source investigated by this study. 
The ratio of the injected energy to the initial kinetic energy governs how much fraction of the ejecta is influenced by the hot bubble expansion (vertical axis). 
On the other hand, the ratio of the injection timescale to the photon diffusion timescale governs the radiative efficiency (horizontal axis). 
The expected observational features and the corresponding classes of energetic SNe are also summarized in each panel. }
\label{fig:engine_driven_sne}
\end{center}
\end{figure*}

As some previous studies \citep[e.g.,][]{2015MNRAS.454.3311M,2016ApJ...818...94K,2018MNRAS.481.2407M} have already suggested, SN ejecta harboring a central power source with different sets of the energy and the timescale potentially explain the diversity of energetic SNe and other high-energy astrophysical phenomena, such as GRBs. 
Figure \ref{fig:engine_driven_sne} presents the landscape of central engine-powered SNe in $E_\mathrm{in}/E_\mathrm{sn}$--$t_\mathrm{in}/t_\mathrm{diff}$ plot implied by our findings. 
Although the simulation results suggest that the ratios $E_\mathrm{in}/E_\mathrm{sn}$ and $t_\mathrm{in}/t_\mathrm{diff}$ are important measures of the impact of the energy injection into the SN ejecta, an apparent caveat is that the initial kinetic energy $E_\mathrm{sn}$ and the diffusion timescale $t_\mathrm{diff}$ of the SN ejecta before the energy injection are not easily known by simply observing individual SLSNe-I. 
The energy injection accelerates the ejecta and modifies the ejecta structure, erasing the information on the kinetic energy and the diffusion timescale of the original ejecta. 
Therefore, the classification in Figure \ref{fig:engine_driven_sne} based on $E_\mathrm{in}/E_\mathrm{sn}$ and $t_\mathrm{in}/t_\mathrm{diff}$ and the following discussion should be considered as rough guidelines.

When the injection timescale is much shorter than the photon diffusion timescale (the left region in Figure \ref{fig:engine_driven_sne}), $t_\mathrm{in}\ll t_\mathrm{diff}$, most of the energy would have been injected before the ejecta starts emitting thermal photons efficiently. 
The injected energy is immediately thermalized due to the high density in the ejecta, and it is subsequently lost by adiabatic cooling. 
As a result, the injected energy ends up as the kinetic energy of the expanding ejecta. 
In this case, an injected energy smaller than the initial kinetic energy, $E_\mathrm{in}/E_\mathrm{sn}\lesssim 1$, leads to a hot bubble stuck in the inner ejecta, leaving outer ejecta unaffected. 
This would make little difference from stripped-envelope SNe with normal explosion energies, although the internal mixing is expected to some extent. 
An injected energy larger than the initial kinetic energy by a factor of a few or more, $E_\mathrm{in}/E_\mathrm{sn}\gtrsim 2$--$3$, the forward shock driven by the hot bubble penetrates the outer ejecta, accelerating the outer layers to relativistic speeds. 
This hot bubble breakout realizes energetic ejecta with a relatively flat radial density profile with the well mixed chemical composition. 
In terms of the optical spectrum, such objects would be seen as an SN Ic-BL. 
The recent light curve analysis of SNe Ic-BL found by iPTF \citep{2019A&A...621A..71T} suggests that a considerable fraction of their sample require almost uniform nickel mixing, which is in agreement with this picture. 
The blast wave driven by the energetic ejecta produces synchrotron and inverse Compton emission. 
Therefore, they are also seen as bright radio sources, such as relativistic SNe like SN 2009bb and 2012ap.

An injection timescale approaching the diffusion timescale, $t_\mathrm{in}/ t_\mathrm{diff}\simeq0.01$--$0.1$, leads to a high conversion efficiency of the injected energy into thermal radiation (the right region in Figure \ref{fig:engine_driven_sne}). 
The bright thermal emission potentially explains the extreme brightness of SLSNe-I. 
The spectral evolutions of such objects would depend on the ratio of the injected energy to the initial kinetic energy. 
An injected energy only comparable to the initial kinetic energy, $E_\mathrm{in}/E_\mathrm{sn}\simeq 1$, again has a limited impact on the ejecta structure. 
The hot bubble is stuck in the inner ejecta and the outer ejecta is left well-stratified. 
The transport of the injected energy is realized mostly by photon diffusion throughout the ejecta. 
Therefore, the photosphere stays in the well-stratified layers. 
The continuous heating of the outer layers can make the ionization structure different from normal SNe.

An injected energy larger than the initial kinetic energy by a factor of a few or more, $E_\mathrm{in}/E_\mathrm{sn}\gtrsim 2$--$3$, makes the hot bubble expansion violent. 
Almost the entire SN ejecta could be affected by the energy injection. 
The receding photosphere soon reaches the layers with the internal mixing. 
Therefore, the photospheric emission spectra would exhibit a unique spectral evolution. 
After the inner ejecta has been revealed, the spectral evolution would be similar to SNe Ic-BL in their nebular phases, which are characterized by broad emission features.

Although we do not investigate in this paper, there might exist central power sources with injection timescales much longer than the photon diffusion timescale. 
Most energy injected from the central power source would be lost in the form of non-thermal emission, as it is deposited when the SN ejecta have entered the supernova remnant stage. 
In this case, we expect much less influence on the SN ejecta and the associated optical emission.

\section{Conclusions\label{sec:conclusions}}
In this work, we performed 2D radiation-hydrodynamic simulations of SN ejecta with a central power source to investigate how radiation transport affects the dynamical evolution. 
The simulation results highlight its importance. 
The amount of the injected energy compared with the initial kinetic energy of the SN ejecta determines how much fraction of the ejecta is destructed by the hot bubble expansion. 
The injection timescale compared with the photon diffusion timescale in the ejecta controls how much fraction of the injected energy is converted to the thermal radiation energy. 
This finding may explain the diversity of unusual SNe, such as SLSNe and SNe Ic-BL, in a unified scheme. 

The impact of the central energy injection should leave some footprint that can be probed by multi-wavelength follow-up observations of energetic SNe. 
In this work, we summarize the expected observational features; optical light curve variations, spectral evolution in the photospheric and nebular phases, polarization signals, and non-thermal emission, based on the simulation results. 
Although more detailed investigations, such as line transfer calculations based on multi-dimensional ejecta structure, are needed to clarify what exactly SN ejecta powered by a central engine looks like, theoretically suggested observational features combined with observed samples of energetic SNe help identifying a central engine embedded in SN ejecta and its role in the energy deposition and transport in the ejecta.

\acknowledgements
We appreciate the anonymous referee for his/her constructive comments on the manuscript. 
A.S. acknowledges support by Japan Society for the Promotion of Science (JSPS) KAKENHI Grand Number JP19K14770. 
K.M. acknowledges support by JSPS KAKENHI Grant (JP20H00174, JP20H0473, JP18H05223). 
Numerical simulations were carried out by Cray XC50 system operated by Center for Computational Astrophysics, National Astronomical Observatory of Japan.

\software{Matplotlib (v3.2.1; \citealt{2007CSE.....9...90H})
}

\appendix

\section{Non-thermal radiation transport test}\label{sec:gamma_ray}
\begin{figure}
\begin{center}
\includegraphics[scale=0.6]{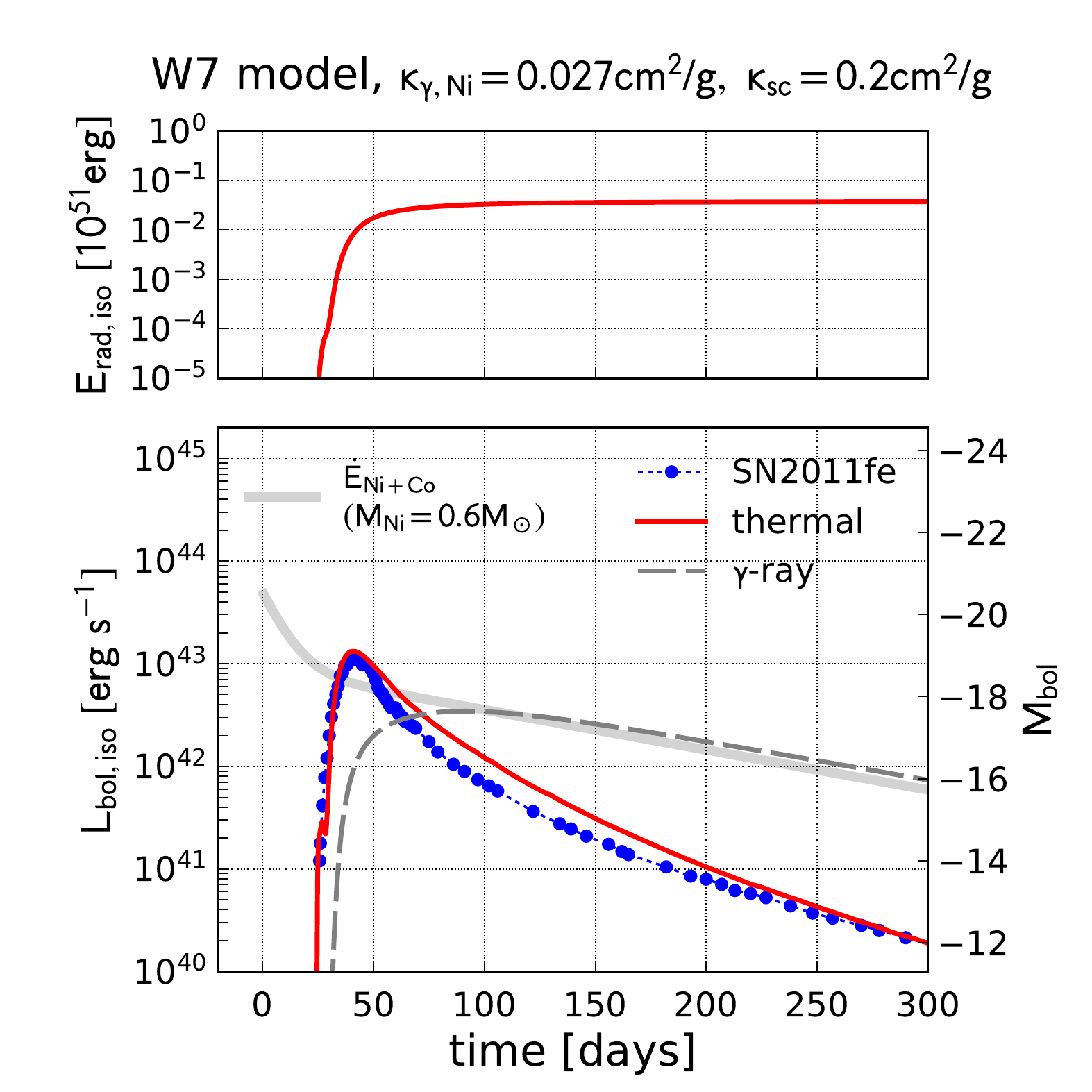}
\caption{Result of the Type-Ia SN light curve test. 
The thermal and non-thermal light curves (lower panel) and the cumulative thermal radiated energy (upper panel) are plotted.  
In the lower panel, we also plotted the energy deposition rate of the nickel decay chain ($M_\mathrm{Ni}=0.6M_\odot$; thick gray line) and the bolometric light curve of Type-Ia SN 2011fe (blue circles). 
}
\label{fig:W7}
\end{center}
\end{figure}

As we have described in Section \ref{sec:radiative_transfer}, the non-thermal radiation component is treated in a separate way from the thermal photons. 
In this section, we provide the results of a test for non-thermal radiation transport. 
We employ the optical and gamma-ray photon transport in the so-called W7 model of Type-Ia SN \citep{1984ApJ...286..644N}, which is one of the theoretical ejecta models best reproducing observational features of Type-Ia SNe. 
The W7 model provides the 1D radial profiles of hydrodynamic variables as well as the mass fractions of chemical elements. 
We carry out a 1D spherical radiation-hydrodynamical simulation by adopting the hydrodynamic profiles provided by the model. 
The model produces $M_\mathrm{Ni}\sim 0.6\ M_\odot$ of radioactive nickel in the inner ejecta. 
The spatial distribution of the nickel mass fraction is treated as a passive scalar in our simulation, which serves as a non-thermal photon source. 
We assume the same absorption and scattering opacity (Equations \ref{eq:kappa_a} and \ref{eq:kappa_es}) as our simulations in the main text. 
The energy injection rate per unit mass is assumed to be
\begin{equation}
    \dot{e}_\mathrm{Ni+Co}=
    \dot{e}_\mathrm{ni}e^{-t/t_\mathrm{ni}}+
    \dot{e}_\mathrm{co}e^{-t/t_\mathrm{no}}
    ,
\end{equation}
with $\dot{e}_\mathrm{ni}=6.45\times 10^{43}$ erg s$^{-1}$, $\dot{e}_\mathrm{co}=1.45\times10^{43}$ erg s$^{-1}$, $t_\mathrm{ni}=8.8$ days, and $t_\mathrm{co}=111.3$ days \citep[e.g.,][]{1994ApJS...92..527N}.  
For the degradation of nuclear gamma-rays accompanied by radioactive nickel and cobalt decays, a gray opacity of $\simeq 0.03Y_\mathrm{e}$ with the electron fraction $Y_\mathrm{e}=0.5$ is commonly adopted and well reproduces SN light curves  \citep[e.g.,][]{1984ApJ...280..282S,1990ApJ...360..242S,1995ApJ...446..766S,1997A&A...328..203C}. 
In this test problem, we assume a non-thermal opacity of $\kappa_\mathrm{\gamma,Ni}=0.027$ cm$^2$ g$^{-1}$. 

In Figure \ref{fig:W7}, we show the resulting thermal and non-thermal light curves and the comparison with the energy deposition rate of the nickel decay chain and the bolometric light curve of the nearby Type-Ia SN 2011fe taken from \cite{2016ApJ...820...67Z}. 
The numerical light curve is in excellent agreement with the observed bolometric light curve around the maximum. 
This reassures that the treatment of the simplified non-thermal photon transfer with an appropriate non-thermal opacity can be used for SNe with a well-known power source. 
At later epochs, the numerical light curve shows a slight deviation from the observed one. 
This could be due to the simplified treatment of non-thermal photons, i.e., the immediate conversion of non-thermal photons to thermal photons. 
In reality, the decay chain is more complicated \citep[e.g.,][]{1988ApJ...325..820A}. 
A $^{56}$Co decay produces nuclear line gamma-rays scattering off electrons or directly emit positrons, which is followed by the energy deposition by the non-thermal electrons/positrons. 
The non-thermal electrons may fly long before the complete thermalization. 
In our treatment, however, the thermalization is assumed to happen instantaneously. 
Nevertheless, the immediate conversion of non-thermal photons to thermal photons is a good approximation while the ejecta is sufficiently opaque as suggested by the good agreement with the observed light curve around the peak. 




\bibliographystyle{aasjournal}
\bibliography{refs}




\end{document}